\documentclass[prb,aps,twocolumn,floatfix,showpacs,showkeys,superscriptaddress]{revtex4-2}

\usepackage{graphicx}
\usepackage{epstopdf}
\usepackage{amssymb}
\usepackage{color}

\usepackage{lipsum}
\usepackage{subfigure}
\usepackage{threeparttable}
\usepackage{booktabs}
\usepackage{longtable}
\usepackage{extarrows}
\usepackage{rotating}
\usepackage{tensor}
\usepackage[all,cmtip]{xy}
\usepackage{mathrsfs}
\usepackage{cancel}
\usepackage{soul}
\usepackage{ulem}
\usepackage{amsmath}
\usepackage{amsthm}
\usepackage{amssymb}
\usepackage{natbib}
\usepackage{graphicx}

\begin{document}

\title{Competing charge density wave phases in YNiC$_2$}

\author{Marta Roman}
\email{marta.roman@pg.edu.pl}
\affiliation{Institute of Solid State Physics, TU Wien, Wiedner Hauptstrasse 8-10, A-1040 Wien, Austria}
\affiliation{Institute of Physics and Applied Computer Science, Faculty of Applied Physics and Mathematics, Gdansk University of Technology,
Narutowicza 11/12, 80-233 Gdansk, Poland}

\author{Simone Di Cataldo}
\affiliation{Institute of Solid State Physics, TU Wien, Wiedner Hauptstrasse 8-10, A-1040 Wien, Austria}
\affiliation{Dipartimento di Fisica, Sapienza University of Rome, Piazzale Aldo Moro 5, 00185 Rome, Italy}

\author{Berthold St{\"o}ger}
\affiliation{X-Ray Center, TU Wien, Getreidemarkt 9, A-1060 Wien, Austria}

\author{Lisa Reisinger}
\affiliation{Institute of Solid State Physics, TU Wien, Wiedner Hauptstrasse 8-10, A-1040 Wien, Austria}

\author{Emilie Morineau}
\affiliation{Institute of Solid State Physics, TU Wien, Wiedner Hauptstrasse 8-10, A-1040 Wien, Austria}

\author{Kamil K. Kolincio}
\affiliation{Faculty of Applied Physics and Mathematics, Gdansk University of Technology,
Narutowicza 11/12, 80-233 Gdansk, Poland}

\author{Herwig Michor}
\email{michor@ifp.tuwien.ac.at}
\affiliation{Institute of Solid State Physics, TU Wien, Wiedner Hauptstrasse 8-10, A-1040 Wien, Austria}

\begin{abstract}
Charge density wave (CDW) orders in YNiC$_2$ are studied by means of combined experimental and computational techniques.
On the experimental side, single crystals grown by the floating-zone method were examined by means of X-ray diffraction, as well as transport and thermal techniques. 
Density functional theory (DFT) calculations founded on the experimentally determined parent and CDW-modified crystal structures provide details of electronic and phononic structures as well as electron-phonon coupling, and resolve changes inflicted upon entering the different CDW phases.
Thereby, contrasting effects of subsequently emerging CDW states characterized by incommensurate $q_{1ic}$ and commensurate $q_{2c}$ modulation vectors are revealed. 
The former state, on-setting below $T_{\text{1ic}}\simeq 305$\,K, weakly modifies the electronic structure by opening an almost isotropic gap on a minor part of the Fermi surface (FS). 
The latter phase, which takes over below $T_{\text{2c}}\simeq 272$\,K  has a more pronounced impact on physical properties via a decomposition of larger parts of the FS. 
These dissimilar behaviors are directly reflected in the electronic transport anisotropy, which is significantly weakened in the $q_{2c}$-type CDW state. 
As revealed by our DFT studies, CDW phases are very close in energy and their origin is directly related to the anisotropy of electron-phonon coupling, which is linked to a specific orbital character of related FS sheets.
Specific heat and thermal expansion studies reveal a nearly reversible first-order phase transition at around $T_{\text{2c}}\simeq 272$\,K, where both CDW phases co-exist within a $T$-interval of about 10\,K.    
\end{abstract}

\date{\today}

\maketitle

\section{Introduction}
Electron-electron and electron-phonon couplings give rise to a broad horizon of physical phenomena, including superconductivity, and spin- or charge density waves. 
The presence of multiple coexisting, intertwined or competing charge density waves (CDW) is typically associated with a geometry of the Fermi surface (FS) that enables multiple electronic instabilities, or with a strongly momentum-dependent electron-phonon coupling \citep{gruner_density_1994, Gruner1988, Monceau2012}. 
While the former, geometric term prevails in systems with an essentially 1D electronic structure, giving rise to nesting-susceptible planar FS sheets, the latter plays a more decisive role in real systems with higher dimensionality \citep{zhu_classification_2015, Eiter_2013, Zhu_2017, Johannes2008}. 

The coexistence or competition of distinct CDW instabilities has been observed in various systems, including transition metal dichalcogenides \citep{Si_2020, Bai2023, Hwang_2024}, bronzes \citep{Foury_2002}, rare-earth tellurides \citep{Malliakas2008, Moore_2010, Hu_2014, Fu_2016}, or recently, in metals with kagome-lattice crystal structure~\citep{Teng2022, Teng2023, Cao2023}.
In extreme cases, when the presence of such multiple degrees of freedom brings two states close in the scale of free energy, a small imbalance, driven by external factors such as light, may become sufficient to open a non-equilibrium state corresponding to another local energy minimum \citep{Kogar2020}.

An abundant sandbox for exploring interactions between various types of order parameters is provided by rare-earth nickel dicarbides $R$NiC$_2$ -- a class of materials known for a rich phase diagram comprising CDW phases, various magnetic states and superconductivity, systematically evolving with the lanthanide contraction \citep{Yamamoto2013,Shimomura2009, Shimomura2016, Roman2018_1, Maeda2019, Roman_2023, Steiner2018}. 
Their electronic structure with quasi-one-dimensional character creates favorable conditions for CDW formation, in some cases with multiple transitions. 
In $R$NiC$_2$ compounds ($R$ = Pr\,--\,Sm) with larger unit-cell volume, CDW superstructures are characterized by a single incommensurate $q_{1ic}=(0.5, 0.5 + \eta, 0)$ modulation wave vector, which may finally lock-in to a commensurate modulation $q_{1c}=(0.5, 0.5, 0)$ at low temperature as reported for GdNiC$_2$ \citep{Yamamoto2013,Shimomura2009, Shimomura2016}. 
Upon replacing the rare-earth with heavier $R$-elements, one observes the emergence and eventually a prevalence of CDW order characterized by a commensurate modulation vector ${q_{2c}}=(0.5, 0.5, 0.5)$ \citep{Maeda2019,Roman_2023, Steiner2018}. 
YNiC$_2$ hosts both, the incommensurate $q_{1ic}$-type and the commensurate $q_{2c}$-type CDW states~\citep{Maeda2019}. 
Despite the observed signatures, the driving forces governing these two electronic instabilities have not been identified and explored yet.

Here we report on single-crystal studies of YNiC$_2$, via X-ray diffraction, thermodynamic and anisotropic transport measurements, aiming to explore the CDW phases and their crossover from the $q_{1ic}$-type CDW state to the $q_{2c}$-type CDW ground state phase. 
Experimental studies are complemented by density functional theory (DFT) calculations of electronic and vibrational properties to investigate the driving forces of CDW formation in this material. 

\section{Experimental}
\label{sect:experimental}
\indent
The polycrystalline material, needed to grow a single crystal, has been synthesized using radio-frequency induction melting in a high-purity argon atmosphere (99.9999\%). 
Pure elements: Y (99.9\%), Ni (99.99\%) metals, and carbon (graphite, 99.999\%) have been used as precursors in this procedure. 
In a next step, polycrystalline feed and seed rods were prepared, using the same induction melting method, and then used to grow monocrystalline YNiC$_2$ via the ﬂoating zone technique in an optical mirror furnace (Crystal Systems Corporation, Japan). 
Single crystals were finally oriented by means of the Laue method and cut along the principal orientations of the orthorhombic parent structure, in dimensions as desired for specific measurements.

An initial characterization of the YNiC$_2$ crystal was performed via scanning electron microscopy (SEM) using a Philips XL30 ESEM with EDAX XL-30 EDX detector and powder X-ray diffraction (pXRD) with an {\it Aeris} powder diffractometer by Malvern Panalytical. 
A homogeneous 1-1-2 stoichiometry with no relevant inclusions of impurities was confirmed by electron microprobe studies and is well supported by pXRD data. 

\begin{table*}
	\center
	\caption{Data collection and refinement details of the incommensurately modulated and triperiodic CDW phases of YNiC$_2$. 
            $R$- and $wR2$-values of the modulated structures at 280 and 250\,K refer to refinements of all reflections, main reflections and $1^{\rm st}$ order satellites, respectively.
             For the commensurate structure at 250\,K, the main reflections are defined as those with even $k$.
             $R_\mathrm{int}$ is given for all reflections (main and satellites).
		 \label{tab:cryst1}
		 }
	\begin{tabular}{llll}
		\hline
		\hline
		$T$ (K)			& 320			& 280 & 250			\\
		$M$			& 171.6				& 171.6 & 171.6		\\
		Space group		& $Amm2$				& $Amm2(\frac12\sigma_20)000$ & $Cm$ 	\\
		$a$ (\AA{})		& 3.5724 (2)				& 3.5699 (5) & 7.5223 (7)\\
		$b$ (\AA{})		& 4.5069 (3)			& 4.5062 (6) & 7.1387 (6)	\\
		$c$ (\AA{})		& 6.0321 (5)		& 6.0267 (9) & 3.7635 (4)		\\
		$\beta$ (deg.)		& 90				& 90 & 106.498 (7)	\\
		$V$ (\AA{}$^3$)		& 97.120 (12)			& 96.95 (2) & 193.78 (3)	\\
		$Z$			& 2				& 2 & 4		 \\
		$D_\textrm{calcd}$ (g cm$^{-1}$) & 5.869				& 5.879 & 5.883		\\
		$\mu$ (mm$^{-1}$)	& 38.907			& 38.975  & 38.999	\\
		Crystal size ($\mu$ m$^3$)	& {$50\times 28\times 5$ \hspace{15mm}}   & {$90\times 77\times 60$ \hspace{5mm}}  & {$60\times 37\times 20$} \\
		$\theta_\textrm{max}$	& 41.21			& 45.28 & 41.09	\\
		Reflections & & & \\
		\quad measured		& 2780				&  {4548} & 8226 		\\
		\quad unique main		& 378				& 467  &  {583}		\\
        \quad\quad  $1^{\rm st}$ order satellites & -- & 794 &  {1073}   \\
		{\quad observed ($I>2\sigma(I)$) \hspace{8mm}} 			& 377				& 467 &  {543}		\\
        \quad\quad  $1^{\rm st}$ order satellites & -- & 469 &   {414}  \\
		Parameters		& 17					& 22 & 32	\\
		$R_\mathrm{int}$	& 0.0236		& 0.0298 & 0.0524		\\
		$R$ ($I>2\sigma(I)$)	& 0.0159			& {0.0536, 0.0502, 0.0844 \hspace{5mm}} &  {0.0431, 0.0398, 0.0652}	\\
		$wR2$ (all)		& 0.0400			& 0.1379, 0.1275, 0.1813 &  {0.1116, 0.1020, 0.1659}	\\
		GooF			& 1.73				& 1.78 & 1.18		\\
		Extinction (Gaussian)	& 180 (15)			& 160 (40) & 138 (17)	\\
		Diff.~el.~density 	&	& 		&  \\

		min, max (e \AA{}$^{-3}$) & $-0.44$, $0.67$		& $-1.79$, $2.01$ & $-3.62$, $2.70$ \\
		Twin operation		& --				& --  & $2_{[102]}$	\\
		Twin volume fraction	& --			& --	& 73.2:26.8 (4)		\\
		CSD Number		& 2392983		& --			& 2392982 \\
		\hline
	\end{tabular}
\end{table*}

Single crystal diffraction data of small fragments of YNiC$_2$ were collected on a \texttt{STOE Stadivari} diffractometer system \citep{stoe} equipped with a \texttt{Dectris EIGER} CdTe hybrid photon counting detector using Mo$K\overline\alpha$ radiation in a dry stream of nitrogen in the 250 to 300\,K range. 
Data were processed using the \texttt{X-Area} software package and a correction for absorption effects applied using the multi-scan approach implemented in \texttt{LANA}~\citep{stoe}.
The low-temperature (LT) commensurate CDW phase was treated as a twin of index 2 (`HKLF5' style reflection data).
Initial models were generated using the coordinates of the isotypic LuNiC$_2$ structures~\citep{Steiner2018}.
The structures were refined against $F^2$ using \texttt{Jana2006}~\citep{jana}. 
All atoms were refined with anisotropic displacement parameters (ADPs).
In the twinned LT phase the ADPs of the two C and the two distinct Y atoms were constrained to be equal up to the $2_{[001]}$ operation of the high-temperature phase. The correct orientation of all domains was unambiguously established based on the Flack parameter.
Data collection and refinement details are compiled in Table~\ref{tab:cryst1}.
Further details on the crystal structure analyses of the triperiodic structure can be obtained from the
inorganic crystal structure database (ICSD) \citep{icsd} on quoting the depository listed at the end of Table~\ref{tab:cryst1}. 

To study thermal expansion above room temperature, pXRD patterns were collected from 300 to 380\,K using a Panalytical X'Pert Pro diffractometer equipped with an Anton-Paar HTK-1200 chamber with He atmosphere. 
Low-temperature thermal expansion (4.3\,--\,310\,K) was measured by capacitive dilatometry employing a tilted plate geometry~\citep{Rotter1998}. 
The size of the cuboid-like oriented YNiC$_2$  crystal used for heat capacity and LT thermal expansion studies was {$ 1.46\times 2.28\times 1.24\simeq 4.13  $~mm$^3$}.

\begin{table}[b]
	\center
	\caption{
		Atomic coordinates and modulation functions in the $q_{1ic}$ incommensurately modulated CDW phase of YNiC$_2$.
		 \label{tab:incomm}
		 }
	\begin{tabular}{lll}
		\toprule 
		\midrule
		$M$	& $x$	& $0$										 \\
			& $y$	& $-0.00591(7)\sin(2\pi\overline x_4)$						 \\
			& $z$	& $0.38848(3)-0.00219(7)\cos(2\pi\overline x_4)$ 				 \\
		Ni	& $x$	& $\frac12+0.01399(13)\sin(2\pi\overline x_4)$				 \\
			& $y$	& $0$								 \\
			& $z$	& $0.00000(6)$								 \\
		C	& $x$	& $\frac12+0.0049(7)\cos(2\pi\overline x_4)-0.0007(6)\sin(2\pi\overline x_4)$	 \\
			& $y$	& $0.3494(7)$		 \\
			& $z$	& $0.1886(5)$		 \\
		\bottomrule
	\end{tabular}
\end{table}

Heat capacity measurements from 2 to 380\,K were carried out on a commercial Quantum Design, Physical Properties Measurement System (PPMS) employing a relaxation-type method. 
Apiezon-H grease was applied for measurements above 300\,K, while Apiezon-N was used as thermal contact medium at below 310\,K.  

A conventional four-probe technique was used for the electrical resistivity measurements in a PPMS system. 
Thin ($\phi$ =50 $\mu$m) gold wires serving as electrical contacts were spark welded to the polished surface of bar-shaped single crystals with typical dimensions $\sim 3\times 1\times 0.5$~mm$^3$ cut along the principal orthorhombic orientations.

DFT calculations were performed using Quantum ESPRESSO~\citep{Giannozzi_JPCM_2009_qe,Giannozzi_JPCM_2017_qe}. 
We employed optimized norm-conserving Vanderbilt pseudo-potentials \citep{Hamann_PRB_2013_ONCV}, with the Perdew-Burke-Ernzerhof exchange-correlation functional. 
Kohn-Sham wavefunctions were expanded in a plane waves basis set using a kinetic energy cutoff of 80\,Ry. 
For calculations of the orthorhombic and $q_{2c}$-type CDW we employed the experimental crystal structures, while the $q_{1ic}$-type structure was described using a 16-atoms commensurate approximant. 
Due to the varying cell size, we employed different meshes for Brillouin zone integration for the parent, $q_{1c}$-type and $q_{2c}$-type unit cells, as detailed in Sect.~I of the Supplemental Material (SM)~\cite{SM}.
\nocite{Cracknell_Book_IrrepsTables_1979,Aroyo_Bilbao1_2006,Aroyo_Bilbao2_2006,momma2008vesta,Kolincio2019}
Densities of states, band structures, and Fermi surfaces were computed non-self-consistently on finer grids as detailed in SM~\cite{SM} Sect.~I. 
The latter were visualized using \texttt{Fermisurfer}~\cite{kawamura2019fermisurfer}.

Phonon calculations were performed using density functional perturbation theory, as implemented in Quantum ESPRESSO, and interpolated using Wannier functions using EPW \cite{ponce_epw_2016, lee_2023_epw} to obtain the phonon frequencies and linewidths from the real and imaginary part of the selfenergy (for further details including the wannierization windows, see SM~\cite{SM} Sect.~II).

The special points for band structures and phonon dispersions for the parent phase are defined according to Ref.~\cite{Cracknell_Book_IrrepsTables_1979}: $\Gamma = (0.00,0.00,0.00)$; $S = (0.00,0.50,0.00)$; $R = (0.00,0.50,0.50)$; $Z =(0.00,0.00,0.50)$; $T =(-0.50,0.50,0.50)$; $Y = (-0.50,0.50,0.00)$; $\Sigma_{0} = (0.39,0.39,0.00)$; $A_{0} = (0.39,0.39,0.50)$. 
The details on how we obtained the corresponding points for the $q_{1c}$- and $q_{2c}$-type CDW cells are given in SM~\cite{SM} Sect.~II.

\begin{figure}[ht]
\includegraphics[angle=0,width=0.96\columnwidth]{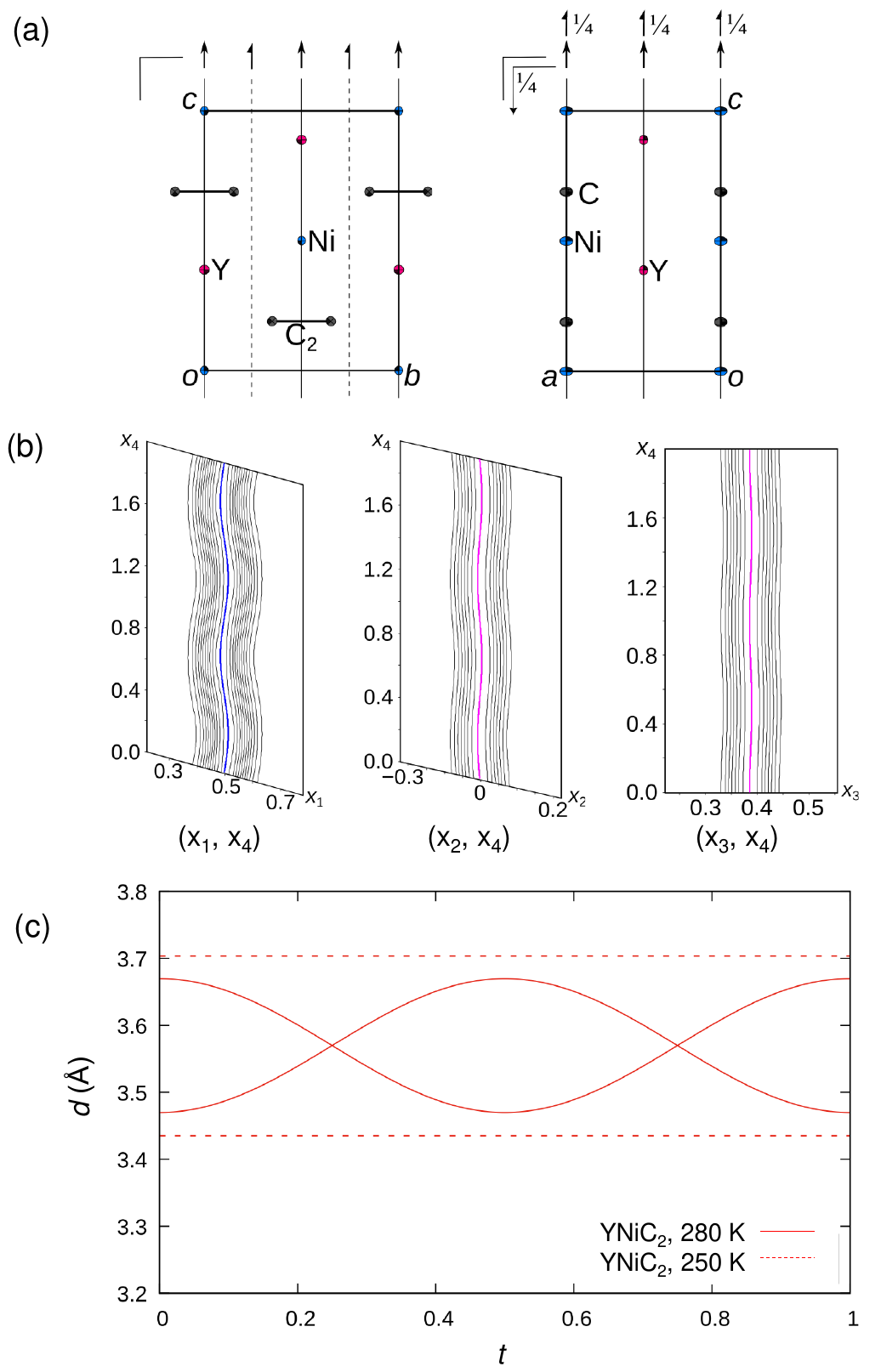}
 \caption{\label{cryst} (a) The crystal structure of the orthorhombic $Amm2$ aristotype phase of YNiC$_2$ viewed along (left) $[100]$ and (right) $[010]$. Y (magenta), Ni (blue) and C (grey) atoms are represented by ellipsoids drawn at the 75\% probability levels. Symmetry elements are indicated using the usual graphical symbols \citep{tablesa-symbols}.
 (b) Sections through the superspace of YNiC$_2$ centered around the (left) Ni and (middle,right) Y atoms in the basic structure. The barycenters of the atoms are indicated by blue (Ni) and magenta (Y) curves. Contours are drawn at the (a) 20 and (b) 50 $e^-$\AA{}$^{-3}$ levels. 
 (c) $t$-plot of the Ni---Ni-distances in $[100]$ direction in the incommensurately modulated phases of YNiC$_2$ red solid line. There are two lines for each case, as each row of Ni atoms extending in the $[100]$ direction alternates between short and long distances (except for $t=0,\frac12$, where Ni atoms are equidistant as in the aristotype structure). For reference, the corresponding distances in the commensurate $Cm$ structures are indicated by dashed lines.}
  \end{figure}

\section{Results and Discussion}

\subsection{Crystal structures}
\label{crst}

\begin{figure*} [ht]

\includegraphics[angle=0,width=1.97\columnwidth]{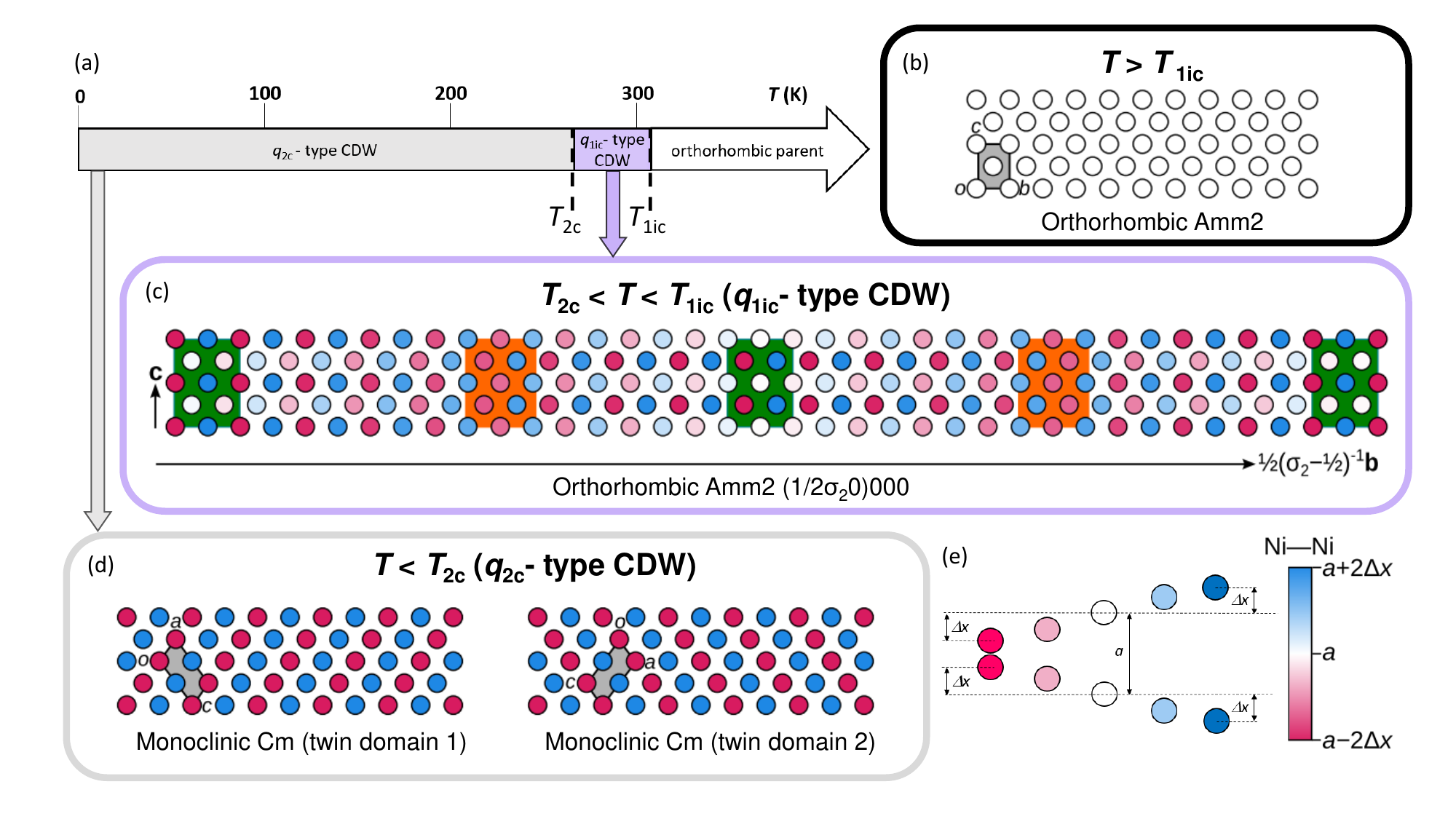}
 \caption{\label{Ni} (a) Phase evolution as function of temperature in YNiC$_2$ compound. (b) - (d) Schematic representation of Ni---Ni-distances in a $(100)$ section of: (b) the $Amm2$ aristotype structure, (c) the incommensurately modulated CDW structure, viewed in $[100]$ direction and (d) both twin domains of the $Cm$ commensurate CDW structures. In panels (b) - (d), the Ni---Ni pairs are represented by disks, and their colors represent the relevant Ni---Ni distance as graphically explained in panel (e); white marks the average value ($=a$ lattice parameter of the parent structure), dark blue and pink colors represent the pairs with longest and shortest Ni---Ni distances, respectively. 
 Y atoms and C$_2$ dumbbells are omitted for clarity.}
\end{figure*}
  
Like all members of the $R$NiC$_2$ family at elevated temperatures, YNiC$_2$ also crystallizes in the orthorhombic $Amm2$ aristotype structure (Fig.~\ref{cryst}a) \cite{Jeitschko1986}. 
All metal atoms and the C$_2$ dumbbells are located on sites with $mm2$ symmetry. 
The Ni atoms and the C$_2$ dumbbells are located on the $x=0$ planes, the Y atoms on the $x=\frac12$ planes.
The Ni and Y atoms form linear chains extending along $[100]$, whereby adjacent atoms are related by the $\mathbf a$ lattice translation with interatomic distances at 320\,K, $a=3.5724$\,(2) \AA{}. 

Figure~\ref{Ni}a sketches the temperature dependent evolution of structural phases. 
Below $T_\mathbf{2c}$, YNiC$_2$ crystallizes as twofold monoclinic $Cm$ superstructure [$\mathbf q_\mathbf{2c}=\frac12(\mathbf a^*+\mathbf b^*+\mathbf c^*)$ modulation of the orthorhombic parent structure], which is isotypic with the corresponding LuNiC$_2$ and TmNiC$_2$ structures. 
The lost point group operations are retained as twin operations (see Table~\ref{tab:cryst1}). 
The structure is characterized by formation of Ni---Ni Peierls pairs in rods extending along $[100]$ with short distances of 3.4351\,(9) \AA{} which alternate with long distances of 3.7036\,(9) \AA{} at 250\,K. 
A detailed description of the monoclinic structure type has been given previously \citep{Steiner2018, Roman_2023}.

Between $T_\textbf{1ic}$ and $T_\mathbf{2c}$, YNiC$_2$ adopts a 1D (along the orthorhombic $b$-axis) incommensurately modulated structure analogous to SmNiC$_2$ \citep{Wolfel2010} with superspace group symmetry $Amm2(\frac12\sigma_20)000$, and modulation wave vector $\mathbf q_{1ic}=\frac12\mathbf a^*+\sigma_2\mathbf b^*$ with $\sigma_2=0.5138(8)$ at 280 K. 
Accordingly, periodicity in the $[100]$ direction is halved (doubling of the $a$-axis, as in the monoclinic $Cm$ structures) fully retained in the $[001]$ direction and lost in $[010]$ direction.
For single crystal XRD data taken right at 275\,K, {\emph i.e.}\ in the vicinity of $T_{2c}$, satellite reflections related to both modulation wave vectors $q_{2c}$ and $q_{1ic}$ are observed  and attributed to a coexistence of $q_{2c}$- and $q_{1ic}$-type CDW domains (see Sect.~\ref{thermodyn} for a discussion on details of the related phase transition and Fig.~S1 in SM~\cite{SM}, for the diffraction image). 
Observe that the $q_{2c}$- and $q_{1ic}$ modulation wave vectors possess fundamentally different $\mathbf c$ components ($\frac12$ and 0, respectively) and therefore the structures cannot be related by a unified superspace description. 
Thus, technically, the $q_{1ic}$ to $q_{2c}$ transition is not of lock-in type.

Since the basic structure of the incommensurate phase is isotypic to the aristotype phase, comparable atomic coordinates were used, up to the $x$-coordinates, which were shifted by $\frac12$. 
The $Amm2$ basic structure contains two symmetrically equivalent reflection planes parallel to $(100)$ (see Fig.~\ref{cryst}a) and the origin may be chosen on either of the planes. 
In the $Amm2(\frac12\sigma_20)000$ superspace group, these reflection planes split in two distinct symmetry elements. 
The $x=0$ plane remains a proper mirror plane in the modulated structure. 
In contrast, reflecting at $x=\frac12$ shifts the modulation functions by half a period, which can be regarded as the incommensurate equivalent of a glide reflection. 
If the $x$-coordinates of the basic structure were retained (Ni at $x=0$), the Ni atoms could not be displaced in the $[100]$ direction. 
The resulting model gives refinements with excellent reliability factors, but is unlikely, given the Peierls pair formation in the commensurate $Cm$ structures.
Note that an analogous phenomenon is observed in the commensurate LT phase: 
The reflection planes parallel to $(100)$ of the aristotype structure split into an alternation of proper reflection planes and glide reflection planes with an intrinsic translation component of $\frac12(\mathbf b+\mathbf c)$ \citep{Roman_2023}. 
There as well, the Ni atoms had to be moved to $x=\frac12$.

The displacement modulation of all atoms was described by first-order harmonics (see Table \ref{tab:incomm}). 
No discontinuity was apparent in the modulation functions (Fig.~\ref{cryst}b). 
Since the Y atoms are located on the $x=0$ reflection plane, displacement is only possible parallel to the $(100)$ plane. 
For symmetry constraints, first-order harmonic modulation waves only allow modulation in the $[100]$ direction for the Ni atoms and the C atoms. 
Theoretically, second-order harmonics could allow for an additional displacement in the $(100)$ plane, but these components refined to zero within experimental error.

The following discussion of the modulation will focus on the the Ni atoms, which are the crucial point of the CDW formation.
The position of the Ni atom located at $\mathbf {\overline x}$ in the basic structure is given as
\begin{align}
	x_1=\mathbf {\overline x}+[\Delta x\sin(2\pi\mathbf {\overline x}\cdot\mathbf q)]\mathbf a,
	\label{eq:1}
\end{align}
where $\Delta x$ is the sine-component of the Ni displacement modulation wave listed in Table \ref{tab:incomm} and assuming that the modulation phase $t$ equals 0 for brevity. 
Since the $\sigma_1$-component of the $\mathbf q$-vector is $\frac 12$, the Ni atom translated by an $\mathbf a$ lattice translation in the basic structure is located at
\begin{eqnarray}\label{eq:2}
	x_2 & = & \mathbf {\overline x}+\mathbf a+\{\Delta x\sin[2\pi(\mathbf {\overline x}\cdot\mathbf q+\frac 12)]\}\mathbf a \\\nonumber
	 & = & \mathbf {\overline x}+[1-\Delta x\sin(2\pi\mathbf {\overline x}\cdot\mathbf q)]\mathbf a
\end{eqnarray}
By subtraction of Eqns.~\ref{eq:1} and \ref{eq:2}, one notes that in the $[100]$ direction Ni atoms are arranged in lines spaced alternately by
\begin{align}
	x_1-x_2=[1\pm2\Delta x\sin(2\pi\mathbf {\overline x}\cdot\mathbf q)]a,
     \label{eq:3}
\end{align}
which corresponds to the Peierls pairs also observed in the commensurate $Cm$ phases.
Here however, the amplitude of the Peierls pair formation varies across the structure.
Since the modulation functions are continuous, each Ni---Ni-distance in the $a-2\Delta x$ to $a+2\Delta x$ range is realized. In particular, rods where the Ni---Ni distances are $a$ correspond to the $Amm2$ aristotype structure with equidistant Ni atoms.

\begin{figure*}[ht]
\includegraphics[angle=0,width=1.86\columnwidth]{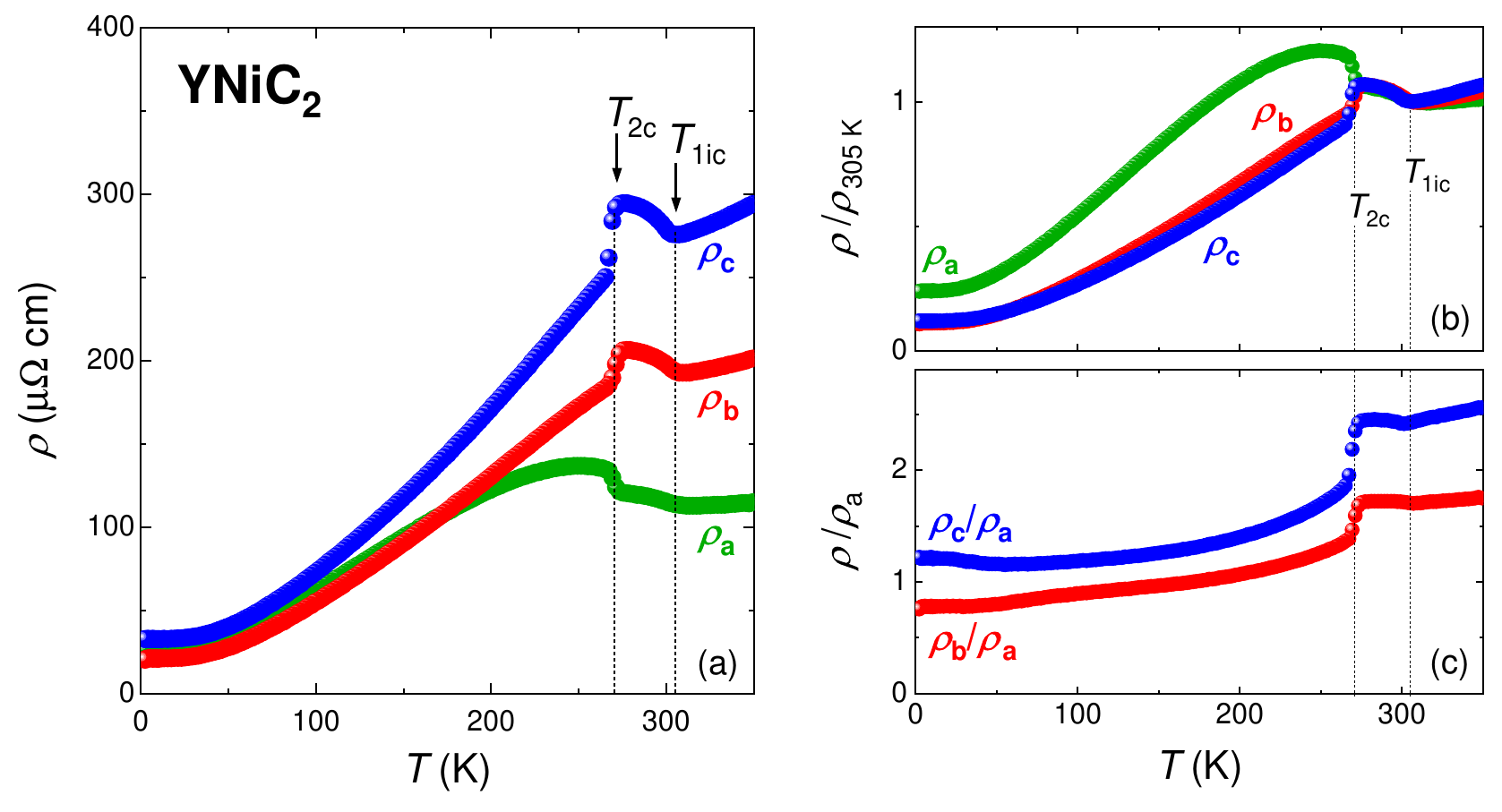}
 \caption{\label{resfig} (a) Temperature-dependent electrical resistivity, $\rho(T)$, measured on bar-shape YNiC$_2$ single crystals which were cut along orthorhombic $a$-, $b$- and $c$-orientations; (b) normalized electrical resistivity $\rho(T)/\rho(T_{1ic})$; (c) temperature-dependent relative anisotropy of the electrical resistivity ${\rho_b}/{\rho_a}$ and ${\rho_c}/{\rho_a}$.}
  \end{figure*}

Figure~\ref{cryst}c shows a $t$-plot of the Ni---Ni distances in YNiC$_2$. 
At its maximum, Peierls pair formation is slightly less pronounced than in the commensurate LT phase.
A translation by $\mathbf b$ in the basic structure corresponds to a $t$-shift by $\sigma_2\approx0.514$, \emph{i.e.} short Ni---Ni distances follow long Ni---Ni distances and vice-versa. Moreover, intermediate Ni---Ni distances are followed by intermediate Ni---Ni distances. 
Figure~\ref{Ni}b--d shows schematic representations of structural phases with color-coded Ni---Ni distances in a plane parallel to $(100)$.
The color-code is graphically explained in Fig.~\ref{Ni}e.
The small deviation of $\sigma_2$ from $\frac12$ results in a long modulation wave. 
Since the $t$-plot in Fig.~\ref{cryst}c is symmetric by $t^\prime=t+\frac12$, the modulation wave repeats after $2/0.014\approx36$ unit cells of the basic structure (see horizontal arrow in Fig.~\ref{Ni}c).

The lattice translation $\frac12(\mathbf b+\mathbf c)$ of the $Amm2$ basic structure increases $t$ by $\frac14+\frac{\sigma_2}{2}\approx\frac14$.
Thus, rods with pronounced Peierls pairs are interleaved by rods with virtually no Peierls pair formation (green background in Fig.~\ref{Ni}c, $t\approx\frac{n}{4}$, $n\in\mathbb Z$ in Fig.~\ref{cryst}c). 
In contrast, rods with medium Peierls pairs combine with other rods of such Peierls pairs (orange background in Fig.~\ref{Ni}c, $t\approx\frac{2n+1}{8}$, $n\in\mathbb Z$ in Fig.~\ref{cryst}c).
A putative lock-in phase with $\mathbf q=\frac12(\mathbf a^*+\mathbf b^*)$ could be either of these two extremes.

\subsection{Electrical resistivity studies}
\label{res}

A charge density wave is essentially a coupled lattice and electronic instability, typically associated with the opening of one or more gaps at the Fermi energy. 
The evolution of the electronic component is directly reflected in the charge transport properties, which can be explored by macroscopic observables, such as the temperature-dependent electrical resistivity, $\rho(T)$. 
The latter, {\emph i.e.}\ $\rho_{a}$, $\rho_{b}$, $\rho_{c}$, measured for single crystalline YNiC$_2$ with current applied along principal orientations is displayed in Fig.~\ref{resfig}a and reveals distinct anomalies related to the two CDW transitions discussed in Sect.~\ref{crst}. 
The first transition at $T_{1ic}=305$ K (i) from the orthorhombic CeNiC$_2$-type parent structure to the orthorhombic incommensurate $q_{1ic}$-type CDW and the second transition at $T_{2c}=272$ K (ii) into a monoclinic commensurate $q_{2c}$-type CDW phase are well reflected by characteristic anomalies of the electrical resistivity, kink-like at $T_{1ic}$ and jump-like at $T_{2c}$, respectively.

At temperatures above $T_{1ic}$, $\rho(T)$ data of YNiC$_2$ in Fig.~\ref{resfig}a share the features of an anisotropic electrical resistivity $\rho_c>\rho_b>\rho_a$ as earlier reported for other $R$NiC$_2$ compounds in their CeNiC$_2$-type parent state \citep{Shimomura2009, Shimomura2016, Roman_2023, Steiner2018}. 
At transition (i), a clear minimum at $T_{1ic}$, followed by a $\rho(T)$ increase upon lowering the temperature, is observed for all three orthorhombic orientations (compare normalized resistivity data in Fig.~\ref{resfig}b and relative anisotropy data in Fig.~\ref{resfig}c), thus, revealing an almost unchanged anisotropy of $\rho_{a}$ vs.\ $\rho_{b}$ vs.\ $\rho_{c}$ across $T_{1ic}$. 
The latter is a characteristic feature of the second-order phase transition into the $q_{1ic}$-type CDW state as previously reported for other $R$NiC$_2$ ($R =$ Sm, Gd, Tb) representatives exhibiting this specific $q_{1ic}$-type CDW transition with a preserved orthorhombic point group symmetry \citep{Shimomura2009, Shimomura2016}.
 
\begin{figure*}[t]
\includegraphics[width=1.98\columnwidth]{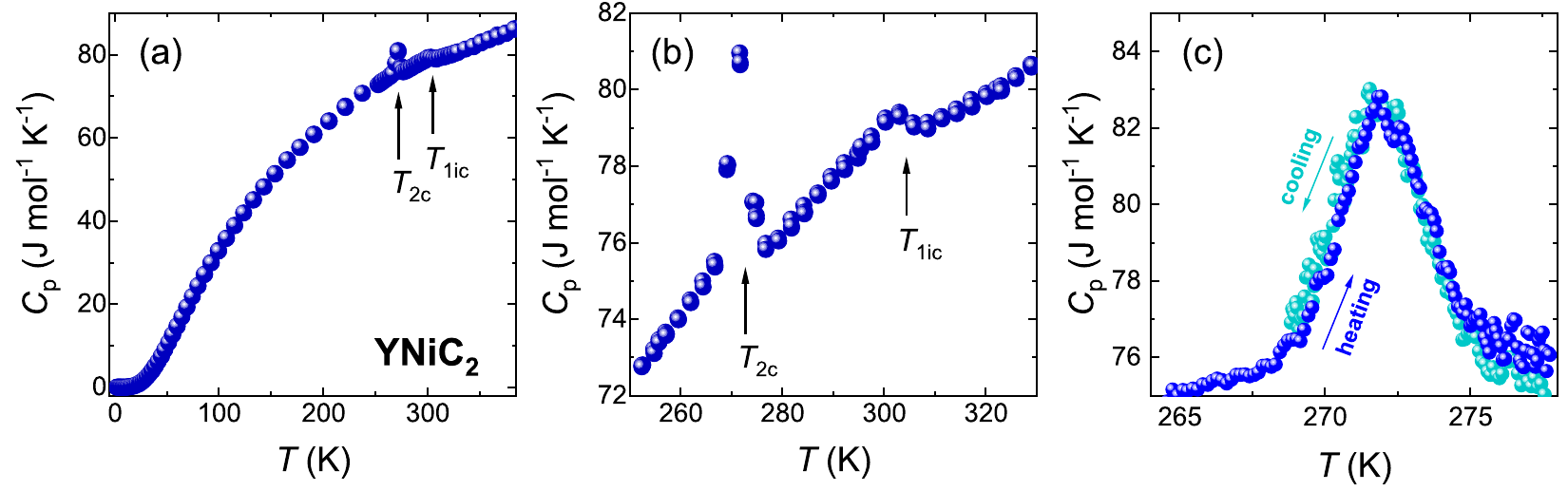}
 \caption{\label{hc} (a) Temperature-dependent specific heat of single-crystalline YNiC$_2$ where CDW phase transitions are marked by black arrows. 
 (b) Expanded view on both CDW phase transitions. 
 (c) Specific heat anomaly related to the first order CDW transition at $T_{2c}$ as determined via a slope analysis of heating (dark blue) and cooling (light blue) curves (see text).}
\end{figure*}

The transition (ii) from the $q_{1ic}$-type into the commensurate $q_{2c}$-type CDW state at $T_{2c}$ is manifested by jump-like anomalies of all three $\rho(T)$ curves in Fig.\,\ref{resfig}a, {\emph i.e.}\ by visible inflections seen simultaneously, yet clearly different for the different orientations. While $\rho_b(T)$ and $\rho_c(T)$ show a sharp downturn, the resistivity measured along the $a$ direction, previously weakly affected at $T_{1ic}$, now visibly increases upon lowering the temperature. 
The pronounced anomaly of $\rho_a(T)$ accompanying the CDW crossover into the $q_{2c}$-type CDW state is not a sole domain of YNiC$_2$, but appears to be a characteristic feature for this type of CDW order.
Similarly specific response of $\rho_a(T)$ has been previously observed for single-crystalline TmNiC$_2$ and LuNiC$_2$ with a single CDW transition directly to the $q_{2c}$-type CDW state \citep{Roman_2023, Steiner2018}. 
With further decrease of temperature, the electrical resistivity gradually decreases for all three orientations. 
Within 100\,--\,200\,K, a visible change of electronic anisotropy takes place with both ${\rho_b}/{\rho_a}$ and ${\rho_c}/{\rho_a}$ ratios gradually decreasing and saturating closer to unity as T is lowered (see Fig.~\ref{resfig}c).
A rather weak anisotropy of the electrical resistivity remains at lowest temperatures, with the lowest value for $\rho_{b}$, similar to earlier reported data of TmNiC$_2$ and LuNiC$_2$~\citep{Roman_2023, Steiner2018}. 
The residual resistivity of YNiC$_2$, as measured at 2\,K, is $\rho_{a0}$ = 27 $\mu\Omega$cm, $\rho_{b0}$ = 20 $\mu\Omega$cm, $\rho_{c0}$ = 33 $\mu\Omega$cm. 

Motivated by a previous transport study of polycrystalline YNiC$_2$, which revealed the presence of a finite thermal hysteresis opening approximately at 285\,K and closing at 265\,K~\citep{Kolincio2019}, we have investigated the character of both CDW anomalies in our single crystal via an electrical resistivity measurement with specific care on an optimal temperature stabilization (see SM~\cite{SM}, Sect.~III for further details).
The corresponding $\rho_c(T)$ measurement (see Fig.~S2 in SM~\cite{SM}) clearly confirms thermal reversibility of the electrical resistivity upon heating and cooling across the second order transition (i) at $T_{1ic}$ and reveals for transition (ii) a very narrow, but nonetheless finite thermal hysteresis of about 0.1\,K for the anomaly at around $T_{2c}$.
From the double-peak-like character of the temperature derivative of $\rho(T)$, we determine the width of transition (ii) ranging from about 265\,to 275\,K (see Fig.~S3a in SM~\cite{SM}). 
Accordingly, the width of the transition ($\sim 10$\,K) reveals to be two orders of magnitude larger than the width of the thermal hysteresis ($\sim 0.1$\,K) and the transition is thus almost perfectly reversible.

\subsection{Specific heat and thermal expansion studies}
\label{thermodyn}

\begin{figure}[hbt]
\includegraphics[angle=0,width=0.98\columnwidth]{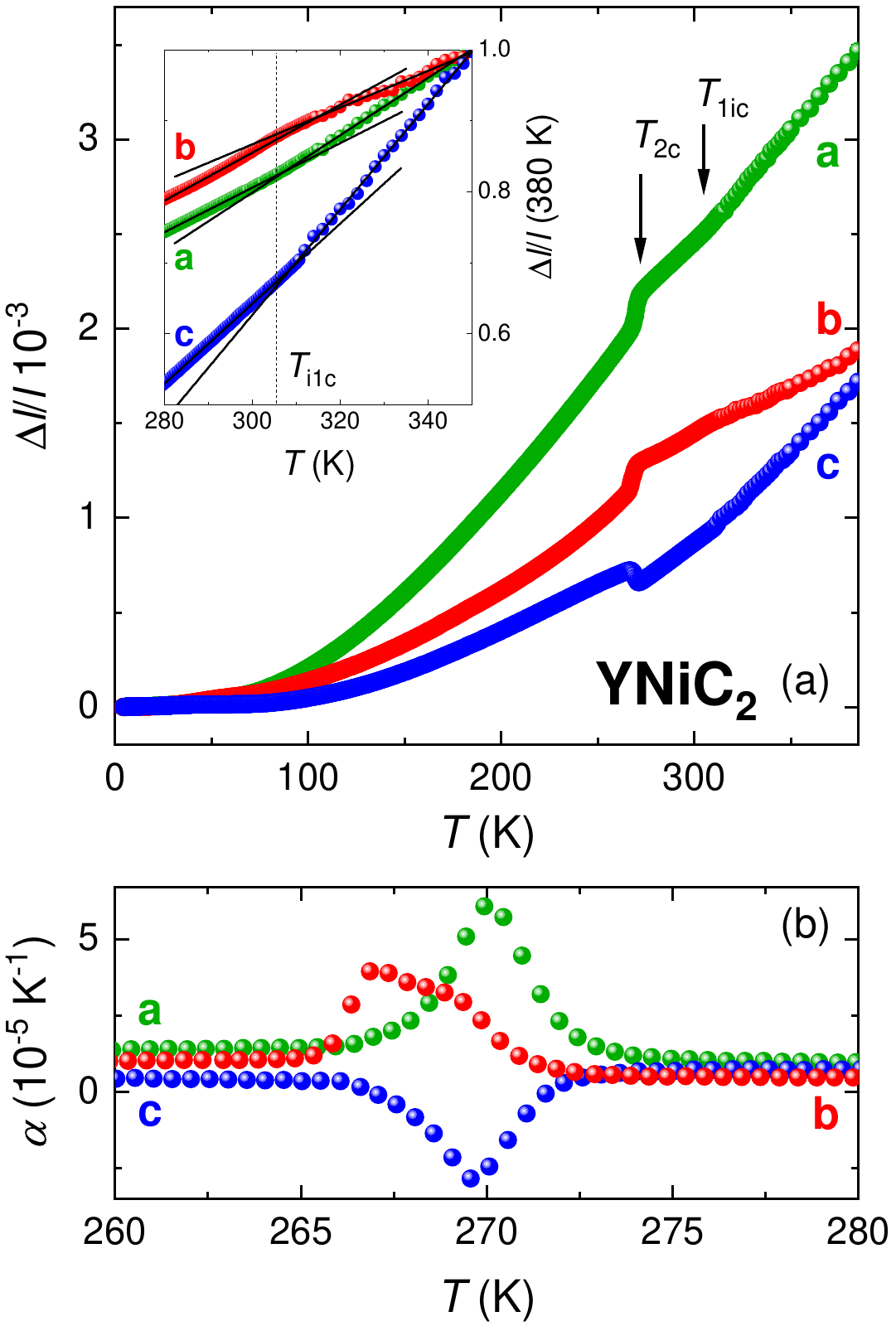}
 \caption{\label{the} (a) Temperature-dependent thermal expansion, $\Delta l/l(T)$, of YNiC$_2$ along orthorhombic $a$-, $b$- and $c$-orientations. 
Inset zooms on corresponding normalized thermal expansion data, $\Delta l(T)/\Delta l(380 {\rm K})$, across the $q_{1ic}$-type CDW transition. 
Solid lines are guides to the eye for revealing slope changes across $T_{1ic}$.  
(b) Expanded view of the thermal expansion coefficients at the $q_{2c}$-type CDW ordering temperature $T_{2c}$.}
  \end{figure}

Closer insight onto the nature of CDW phase transitions in YNiC$_2$ is revealed by specific heat and thermal expansion studies of YNiC$_2$ single crystals. 
The two CDW phase transitions identified by the above XRD and electrical resistivity studies are well reflected by corresponding anomalies of the temperature-dependent specific heat, ${C_p}(T)$, and thermal expansion, $\Delta l/l(T)$, measurements presented in Figs.~\ref{hc} and \ref{the}, respectively. 
Most striking, apart from specific characteristics of the CDW transitions, is the overall strongly anisotropic behavior of the thermal expansion in Fig.~\ref{the}a, which exhibits the largest expansion along the conventional orthorhombic $a$- and the smallest along the $c$-axis, {\emph i.e.}\ YNiC$_2$ shows qualitatively a very similar behavior as earlier reported for TmNiC$_2$ \cite{Roman_2023}. 

Focusing on the CDW phase transitions, a close-up view on the specific heat anomalies presented in Fig.~\ref{hc}b reveals transition (i) from the orthorhombic parent-type to the $q_{1ic}$-type CDW state at $T_{1ic}=305$\,K via a second order jump-like anomaly, $\Delta{C_p} \simeq 1.5(2)$ J\,mol$^{-1}$K$^{-1}$,  and (ii) a first-order phase transition (FOT) from the $q_{1ic}$- to the $q_{2c}$-type CDW state at around $T_{2c} \simeq 272$\,K is indicated by a peak-like anomaly.
Considering the first order nature of the latter transition (ii) and associated latent heat, a standard evaluation of heat capacity data collected with a relaxation-type technique (as applied for Fig.~\ref{hc}b may yield erroneous results and, thus, an adapted slope analysis for heating and cooling curves across that specific heat anomaly has been applied as implemented for this specific purpose in the software of the PPMS heat capacity option). 
Specific heat data resulting from the slope analysis of one such heating-cooling cycle is displayed in Fig.~\ref{hc}c and reveals (within error bars imposed by thermal gradients) an essentially reversible nature of the nearly symmetric peak-shape anomaly when comparing the heating (dark blue) and cooling (light blue) data (results of additional cycles with varying initial temperatures are shown in Fig.~S3b in SM~\cite{SM}) with a very similar width ($\sim 266$\,--\,276\,K) as indicated by the above discussed electrical resistivity data. 
The observed thermal reversibility across this transition is, within error bar, consistent with the above mentioned almost perfectly reversible behavior of the electrical resistivity, but nonetheless unexpected in context with a first order nature of that phase transition. 
A quantitative evaluation of the FOT anomaly in Fig.~\ref{hc}c yields associated enthalpy and entropy changes, $\Delta H\sim 30$\,(3) J/mol and $\Delta S\sim 0.11$(1) J/mol$\cdot $K, respectively.  

Thermal expansion data of YNiC$_2$, $\Delta l/l(T)$ as merged in Fig.~\ref{the}a with data from capacitive dilatometry ($T<310$\,K) and pXRD ($T>310$\,K), feature the CDW phase transitions: (i) via changes of slopes at $T_{1ic}$, {\emph i.e.}\ by the typical thermal expansion feature of a second order phase transition, and (ii) the transition at $T_{2c}$, via a pronounced step-like anomaly of $\Delta l/l(T)$, which is indicative of a first order transition. 
 
For both CDW transitions, the observed orientation dependent changes of the thermal expansion related to these phase transitions, when comparing signs of $\Delta l/l(T)$ anomalies along different orthorhombic orientations, imply reversed thermal shifts of the CDW phase transition upon stress or strain along these orientations (see below). 
At the second order CDW transition (i) at $T_{1ic}$, $\Delta l/l$ curves display rather subtle kink-like anomalies with an increase of slope upon increasing temperature for $a$- and $c$-axis data, but a decrease of slope for $b$-axis data (see inset in Fig.~\ref{the}a). 
Experimental uncertainties caused by merging two different data sets right near transition (i) at $T_{1ic}$ impede a quantitative evaluation of the changes of thermal expansion coefficients $\Delta\alpha$.   
At the first order CDW transition (ii) at $T_{2c}$, jump-like anomalies $\Delta l_a/l_a\simeq + 1.5\times 10^{-4}$ and $\Delta l_b/l_b\simeq + 1.2\times 10^{-4}$ are observed for orthorhombic $a$- and $b$-orientations, {\emph i.e.}\ an expansion when raising temperature across the FOT, but on the contrary a length contraction, $\Delta l_c/l_c\simeq - 1.0\times 10^{-4}$, which add up to a relative volume expansion, $\Delta V/V\sim + 1.7\times 10^{-4}$.

The initial response of the phase transition temperature $T_{2c}$ to uni-axial pressure on YNiC$_2$ (assuming a constrained cross-section for the sake of simplicity) is evaluated via an adapted Clausius-Clapeyron relation, ${{\rm d}T_{2c}}/{{\rm d}\sigma_i} = {(\Delta l/l_i)}V_m/\Delta S\sim \Delta l/l_i\cdot 2.7\times 10^{-4}$ K$\cdot$m$^2$/N, where $\sigma_i$ is an uni-axial pressure along orientation $i$, $V_m=2.92\times 10^{-5}$ m$^3/$mol is the molar volume of YNiC$_2$ and $\Delta S\sim 0.11$(1) J\,mol$^{-1}$K$^{-1}$ is the above determined FOT entropy gain.
The thermodynamic Clausius-Clapeyron relation, thus, yields an increase of $T_{2c}$ upon uni-axial pressure along the orthorhombic $a$- and $b$-axis, {\emph i.e.}\ ${{\rm d}T_{2c}}/{{\rm d}\sigma_a}\sim +4.1\times 10^{-8}$ K$\cdot$m$^2$/N ($+41$ K/GPa) and ${{\rm d}T_{2c}}/{{\rm d}\sigma_b}\sim +3.2 \times 10^{-8}$ K$\cdot$m$^2$/N ($+32$ K/GPa) and, on the contrary, for uni-axial pressure along the orthorhombic $c$-axis a decrease ${{\rm d}T_{2c}}/{{\rm d}\sigma_c}\sim -2.7 \times 10^{-8}$ K$\cdot$m$^2$/N ($-27$ K/GPa). 

In course of the orthorhombic to monoclinic CDW phase transition, the formation of monoclinic twin-domains (see Sect.~\ref{crst}) in combination with the highly anisotropic length changes (see Fig.~\ref{the}a) must obviously result in an evolution of local stress/strain patterns and, thus, in a local variation of $T_{2c}$. 
The latter results in a finite thermal width of the phase transition where $q_{1ic}$- and $q_{2c}$-type CDW domains coexist. 

\begin{figure*}[hbt]
\includegraphics[angle=0,width=2.05\columnwidth]{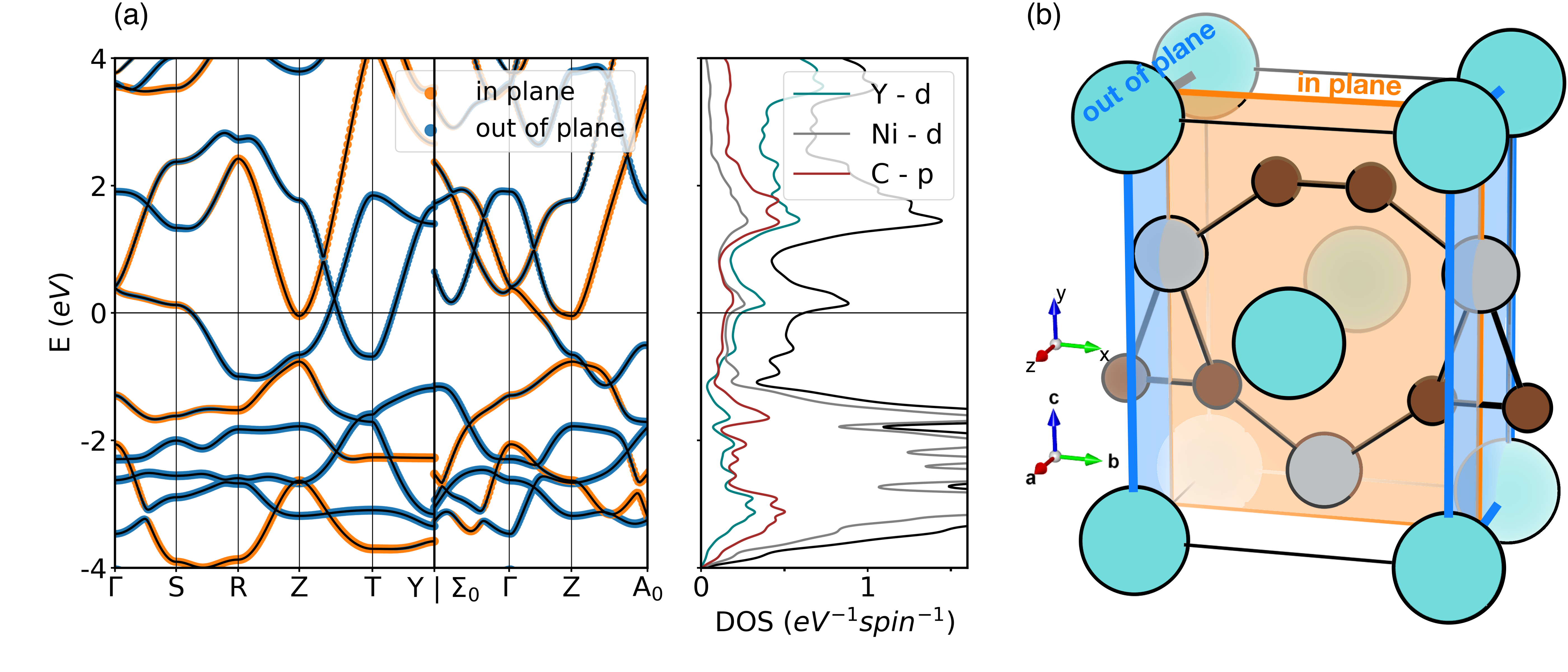}
 \caption{\label{BSorthohor} (a) Electronic band structure and projected DOS in the orthorhombic type parent state of YNiC$_2$. 
The bands are colored with a linear combination of projections onto atomic orbitals. In particular, orange indicates the {Ni-C$_2$} plane corresponding to the \textit{in~plane} combination {of atomic orbitals}, while blue indicates \textit{out~of~plane}, as defined in the text. 
 (b) Sketch of the YNiC$_{2}$ crystal structure highlighting the \textit{in~plane} (orange) and \textit{out of plane} (blue) areas.}
  \end{figure*}

A close-up view in Fig.~\ref{the}b on the anomalies of the thermal expansion coefficients $\alpha_i(T)$ across the phase transition at $T_{2c}$ further reveals an unexpected shape of the anomaly of $\alpha_b(T)$. 
While those of $\alpha_a(T)$ and with reversed sign $\alpha_c(T)$ display a peak-shape similar to that of $C_p(T)$, the  anomaly of $\alpha_b(T)$ exhibits a double peak-like maximum with the absolute maximum being substantially shifted towards the lower onset of the FOT.
In the $q_{2c}$-type CDW state, for the sake of strain-energy minimization, a regular pattern of twin-domains is formed which is accompanied with an accordingly aligned pattern of stress and strain fields. 
With the disappearance of twin-domains across the transition into the $q_{1ic}$-type CDW state all stress and strain related to twin-formation will be relaxed and corresponding strain energy will be released.
In addition, a lattice softening related to CDW sliding degrees of freedom may become relevant at the transition to the incommensurate $q_{1ic}$-type CDW state where translation symmetry is lost along the orthorhombic $b$-axis.
Both aspects may thus contribute to some modifications of the thermal expansion anomalies as compared to the specific anomaly of the transition from the $q_{1ic}$- to $q_{2c}$-type CDW state. 
The strain effects are further relevant with respect to the observed, but unexpected, almost perfect thermally reversible behavior across the FOT at $T_{2c}$.  

A reversible behavior in a first order structural phase transition results from a coupling mechanism which stabilizes specific phase ratios within a finite temperature interval as proposed earlier for a (similar to present case) orthorhombic to monoclinic ferroelectric transition in a hafnium-zirconium oxide film~\cite{Zeng}. 
Local compressive stress evolving from the orthorhombic to monoclinic structure transformation was suggested to stabilize an orthorhombic phase fraction across a wide temperature interval. 
The orthorhombic to monoclinic CDW phase transition of YNiC$_2$, as detailed in the previous paragraph, is well suited to originate local strain patterns which, in course of energy minimization, stabilize a mixed pattern of $q_{1ic}$-type and $q_{2c}$-type CDW domains within a finite temperature interval, {\emph i.e.}\ within the width of the transition. 
A simultaneous appearance of corresponding $q_{1ic}$ and $q_{2c}$ satellite reflections at around 275\,K was also confirmed by single crystal diffraction data discussed in Sect.~\ref{crst}.  

Contrary to the single crystal case, anisotropic thermal expansion of grains in a polycrystal will cause inter-grain strain already upon cooling within the orthorhombic parent state and, indeed, CDW phase transitions of polycrystalline YNiC$_2$ extend over about twice as large temperature intervals (compare Ref.~\citep{Kolincio2019}).
The nature of strain effects, originating from inter-grain strain as predetermined by thermal history in polycrystals versus energy minimizing orthorhombic to monoclinic phase coupling in monocrystals, is rather different and, thus, the suppression of hysteretic behavior expected to be less effective for polycrystals. 
The latter is supported by resistivity data of polycrystalline YNiC$_2$ reported in Ref.~\citep{Kolincio2019} and by a reference heat capacity measurement of polycrystalline YNiC$_2$ analogous to that in Fig.~\ref{hc}c, which however did not at all reproduce the essentially reversible behavior of the single crystal data (compare Figs.~S3b and S3c in SM~\cite{SM}).

\subsection{DFT study of electronic and vibrational properties}
\label{sect:DFT}

To explain the origin of the experimental observations in terms of the electronic structure of YNiC$_2$, we performed \textit{ab initio} DFT calculations on the orthorhombic parent and distorted CDW phases. 
We note that all calculations were performed using the experimental structures, as reported in Sect.~\ref{sect:experimental}, of course, except for the $q_{1ic}$-type CDW phase which lacks translation symmetry along the $b$-axis and is, thus, simplified to a 16-atoms approximant cell, in the following referred to as $q_{1c}$-type CDW. 
For further details we refer the interested reader to SM~\cite{SM}.

Figure~\ref{BSorthohor}a presents the electronic band structure of the orthorhombic CeNiC$_2$ structure-type parent phase of YNiC$_2$, projected onto two different subsets of orbitals, that from here on we will name \textit{in~plane} (orange) and \textit{out~of~plane} (blue). 
Their meaning is sketched in Fig.~\ref{BSorthohor}b. According to this definition, \textit{in~plane} orbitals are C and Ni orbitals lying mainly in the $xy$ plane (orange plane in Fig.~\ref{BSorthohor}b), plus Y orbitals along the $z$ direction. 
Complementarily, \textit{out of plane} orbitals are C and Ni orbitals mainly in the $z$ direction, and Y orbitals in the $xy$ plane (blue planes in Fig.~\ref{BSorthohor}b). 
For a more detailed definition see Ref.~\cite{ip_oop_definition}.

The choice of this unusual partitioning is motivated by the direct observation of the Kohn-Sham wavefunctions near the Fermi energy (Sect.~IV of SM~\cite{SM}), as well as the Wannier functions derived from those bands. 
The bands colored with these projections are shown in Fig.~\ref{BSorthohor}a;  
these are well separated in the \textit{in~plane} (i.e., in the Ni-C plane) and \textit{out~of~plane} characters, with the exception of a single band that changes character as it crosses the Fermi energy in the S--R and $\Gamma$--Z line (coordinates of the reference points of the Brillouin zone are provided in Sect.~\ref{sect:experimental}). 
In addition to the bands, Fig.~\ref{BSorthohor}a shows the atom-projected density of states (DOS) where from -4 to -2 eV, the DOS exhibits mainly a Ni-$d$ character, while in the -2 to 2\,eV window around the Fermi energy the atomic character is evenly spread between Y, Ni, and C, consistent with the hybrid character of these states.

\begin{figure}[t]
\includegraphics[angle=0,width=1.03\columnwidth]{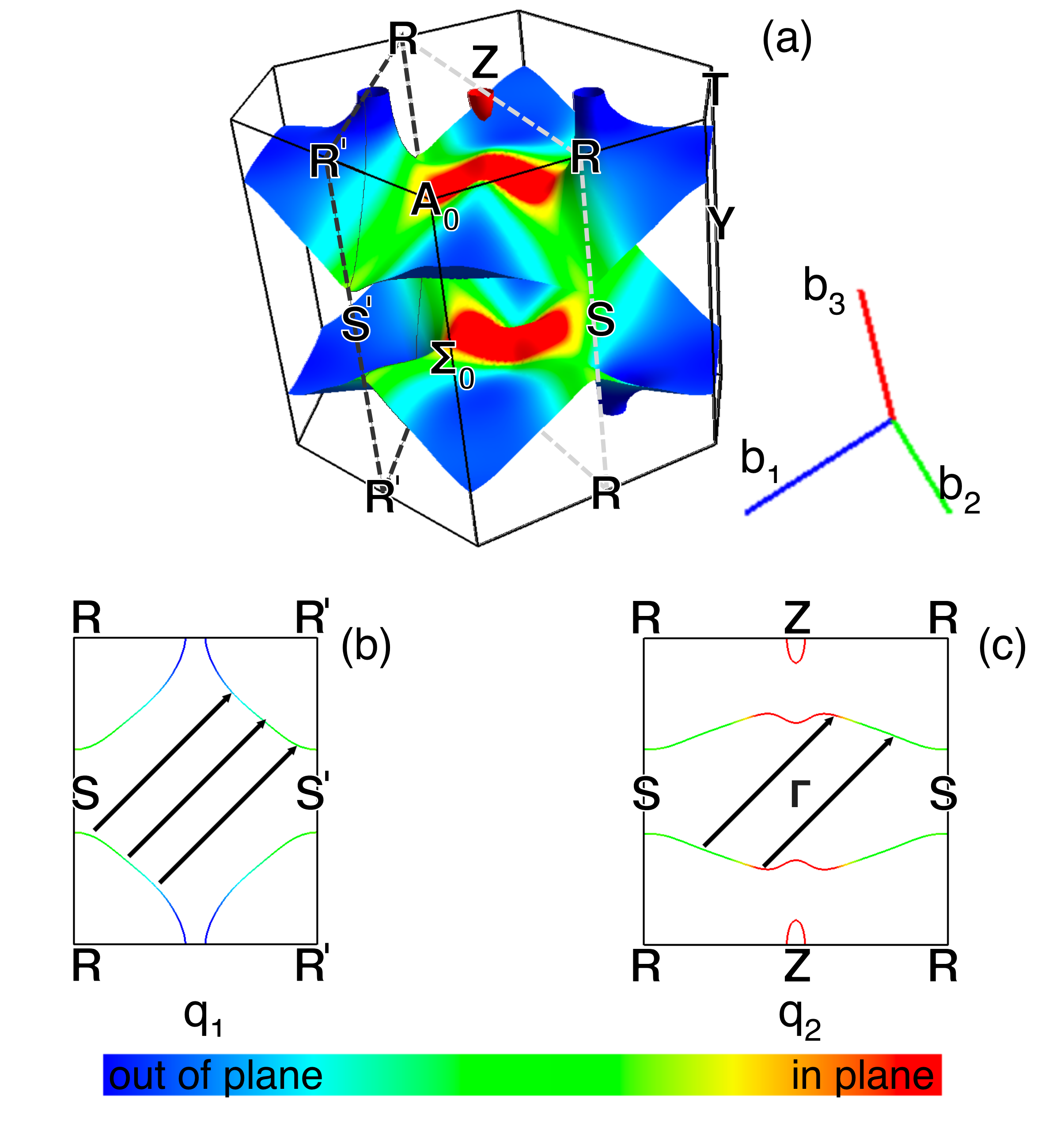}
 \caption{\label{fig7:FS} (a) Fermi surface decorated with an \textit{in~plane} projection, cut with a plane (dashed lines) (b) and (c) 2D cuts indicating nesting vectors with $q_1\sim(0.5, 0.5, 0)$ and $q_2\sim(0.5, 0.5, 0.5)$ momentum, respectively.}
  \end{figure}
  
\begin{figure}[t]
\includegraphics[angle=0,width=0.84\columnwidth]{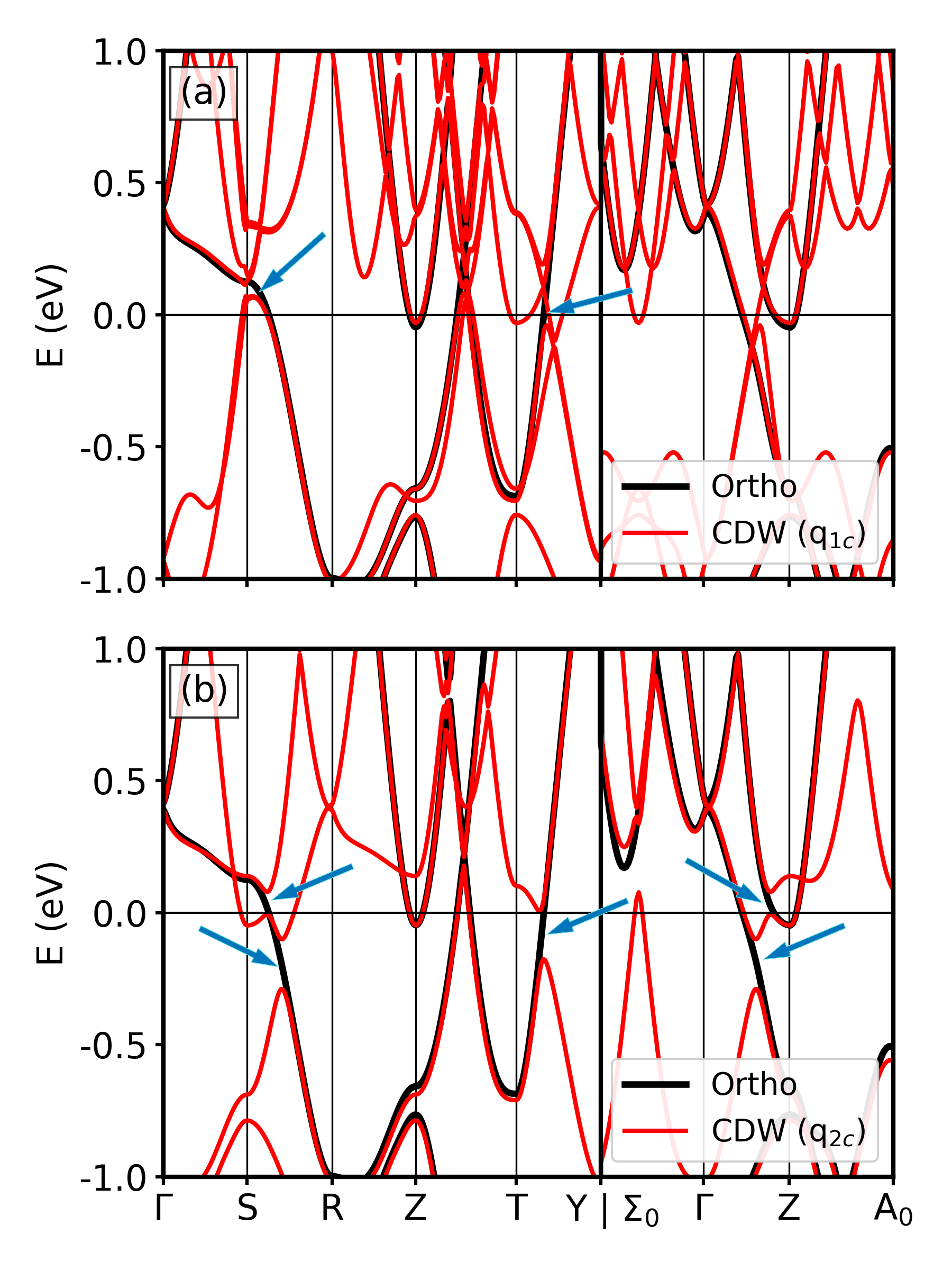}
 \caption{\label{BSCDW} Electronic band structure in $q_{1c}$-type (a) and $q_{2c}$-type (b) CDW state. The black line indicates the orthorhombic parent state, while the red line indicates the respective CDW phase. Blue arrows point at local band splittings.}
\end{figure}

As discussed in the following, the change of band character is closely related to the CDW formation in YNiC$_{2}$. 
The Fermi energy is crossed by bands in the S--R and $\Gamma$--Z lines, and in Z--T and T--Y (see Fig.~\ref{BSorthohor}a).
Resulting FS sheets are shown in Fig.~\ref{fig7:FS}a with a color scheme indicating an \textit{in~plane} projection of the character of electronic states.
There are two different sheets: a main FS sheet formed by open quasi-planar sheets which are connected by hourglass-like pillars, and a small electron pocket centered around the Z point. 
The orbital character of the main FS sheet is extremely uneven (see Fig.~\ref{fig7:FS}), ranging from mainly {\sl in~plane} around the $\Gamma$--Z direction (red color), to mainly {\sl out~of~plane} around Z--T and T--Y (blue color).
It was pointed out earlier, that the strong quasi-planar character of the FS creates proper conditions for nesting with two primary $q$-vectors ({\emph i.e.}\ $q_1\sim(0.5, 0.5, 0)$ and $q_2\sim(0.5, 0.5, 0.5)$) which can be revealed via FS contours obtained at proper 2D cuts of the Brillouin zone \cite{Laverock2009,Kim2013}.
Accordingly, we show in Fig.~\ref{fig7:FS}b and ~\ref{fig7:FS}c two slices of the Brillouin zone hosting $q_{1}$ and $q_{2}$ nesting vectors, which in our {\sl in~plane} projection connect states with \textit{intermediate} (green-colored) character.

In order to investigate the impact of CDW formation on the electronic structure of YNiC$_2$, we directly compare the band structure of the orthorhombic parent phase with that of $q_{1c}$- and $q_{2c}$-type CDW phases in Fig.~\ref{BSCDW}a and \ref{BSCDW}b, respectively, always following the paths in reciprocal space corresponding to the ones selected for the orthorhombic phase so that, folding aside, the bands lie on top of each other (see SM~\cite{SM} Sect.~II\,A for technical details, and Fig.~S6b for the difference in DOS). 
Overall, both CDWs open a pseudogap (defined as a partial depletion of electronic states, where some portions of the Fermi surface develop an energy gap while others remain conductive \cite{timusk1999pseudogap}), and thus, system remains metallic. Here we define the CDW gap as the energy range over which the DOS in the distorted phase is lower than that of the undistorted parent phase (see Fig.~S6). 
In momentum space, the band structure reveals local band splittings, e.g., along the T--Y (25 meV for q$_{1c}$ and 150 meV for q$_{2c}$) and S--R lines, in correspondence to the regions where the orbital character changes from \textit{in~plane} to \textit{out~of~plane}. 
As a result, the only bands that previously changed character are now broken into two separate bands, each with its own well-defined character. 
Although each individual splitting is small in magnitude, the cumulative effect of gap formation at different k-points (and energies) removes electronic states near the Fermi level and is seen as CDW pseudogap in the DOS. 

\begin{figure}[t]
\includegraphics[angle=0,width=0.99\columnwidth]{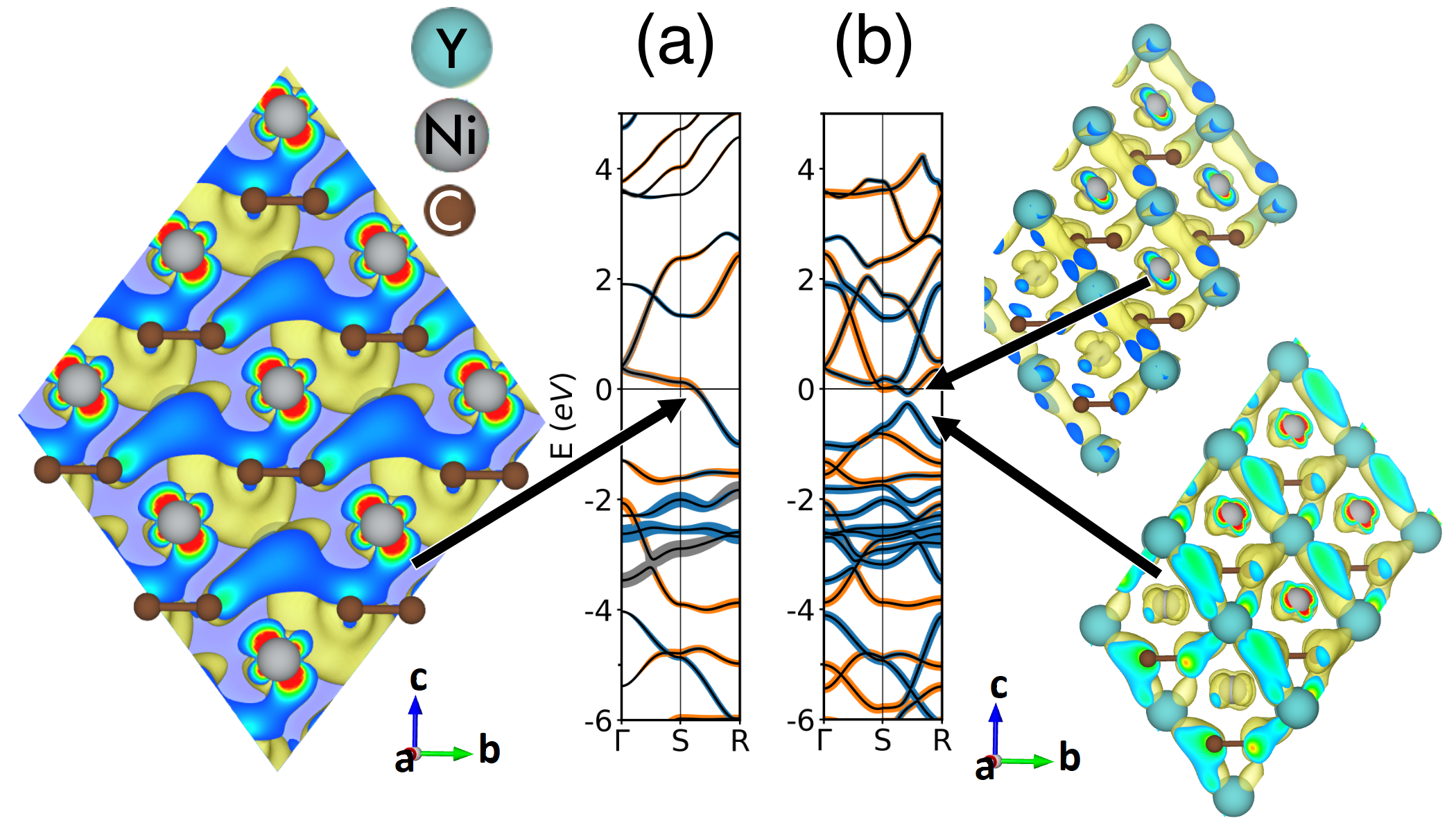}
 \caption{\label{CDW_orbi} Comparison of electronic bands along $\Gamma$--S--R lines, colored with a linear combination of projections onto atomic orbitals, in the orthorhombic parent phase (a) and the $q_{2}$-type CDW phase (b). 
 Aside, we show the probability density from the Kohn-Sham wavefunction indicated with the black arrows with momenta (0.00, 0.50, 0.13) and (-0.047, 0.38, 0.64) for the orthorhombic parent phase and $q_{2}$-type CDW phase, respectively.}
\end{figure}

The separation of wave functions into bonding and anti-bonding states with distinctly distinguished \textit{in~plane} and \textit{out~of~plane} character is highlighted in Fig.~\ref{CDW_orbi} via electron probability density plots as obtained from the Kohn-Sham wavefunction calculated for (a) the orthorhombic parent state and (b) the monoclinic $q_{2c}$-type CDW state (see also SM~\cite{SM}, Fig.~S5 for projected band structures). 
This change of character is directly related to the direction of the atom displacements of the CDWs.
Indeed, both are characterized mainly by the displacement of Ni in the $z$ direction, {\emph i.e.}\ out of the $xy$ planes in which Ni and C atoms are located. 
This changes the hopping integrals, and lowers the energy of bands with \textit{out~of~plane} character, and raises the \textit{in~plane} ones, which lowers the total electronic energy as compared to the undistorted phase.
While both CDWs feature certain similarities, it is clearly noticeable that the electronic structure modification caused by the $q_{2c}$-type CDW is much more pronounced: more, and larger band splittings open in the S--R, T--Y, and $\Gamma$--$Z$ lines mainly below the Fermi energy, while for the $q_{1c}$-type CDW the main splitting is visible slightly above the Fermi level. 
The resulting CDW pseudogap from the DOS analysis is about 283 and 414 meV, for $q_{1c}$- and $q_{2c}$-type CDW, respectively (see Fig.~S6 in SM \cite{SM}).

\begin{figure*}[ht]
\centering
 \includegraphics[width=1.7\columnwidth]{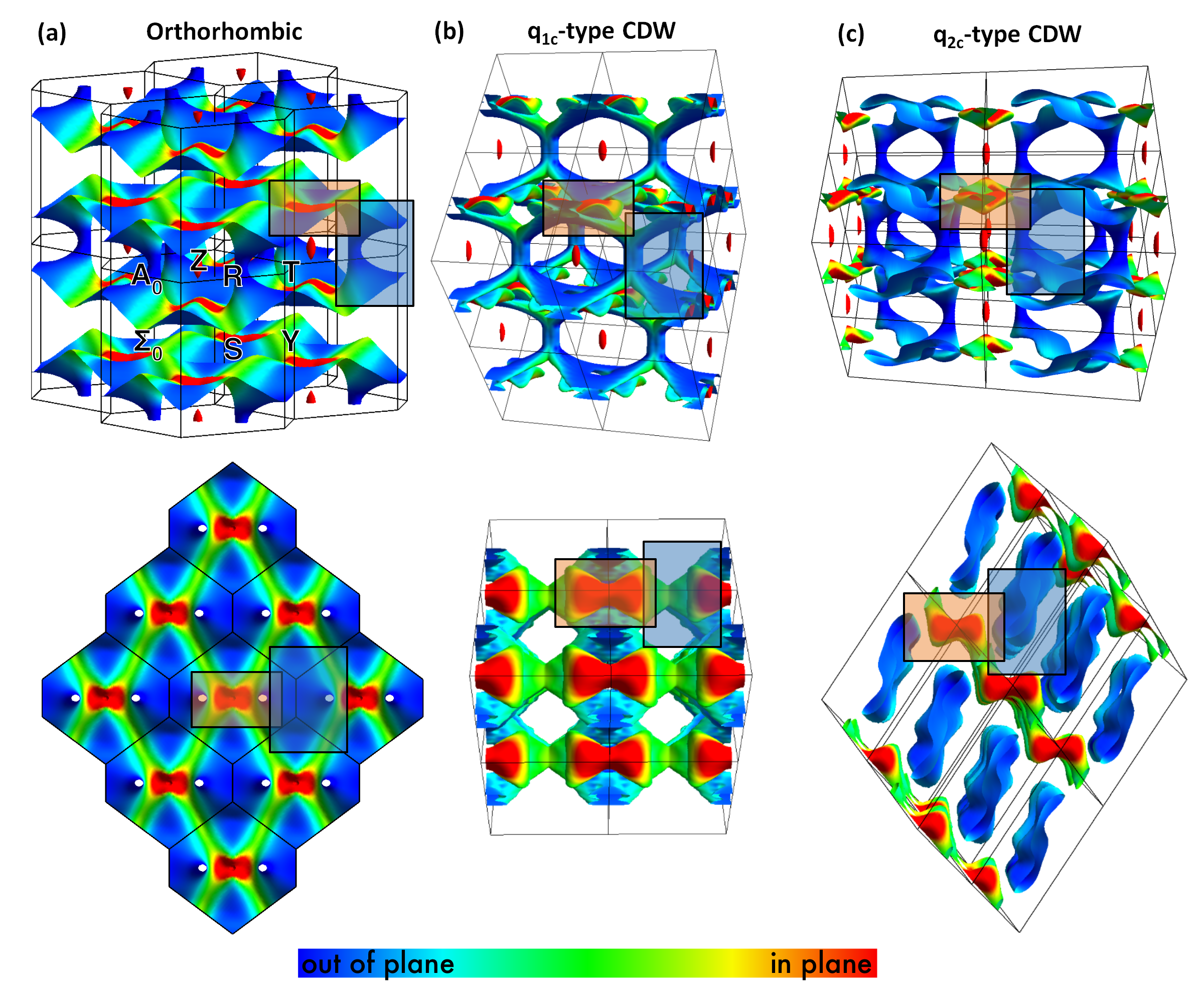}
 \caption{\label{fig:FS} Fermi surface with color-coded projection onto \textit{in~plane} states for (a) orthorhombic parent, (b) $q_{1c}$-type CDW, (c) $q_{2c}$-type CDW states with a side view perspective (top) and top view perspective (bottom). The orange and azure transparent boxes highlight common features of the FS in the different phases: a ribbon-like and an hourglass-like feature, respectively.}
\end{figure*}

Figure~\ref{fig:FS} shows the Fermi surfaces decorated with a projection to \textit{in~plane} states for the orthorhombic parent (a), the $q_{1c}$- (b) and the $q_{2c}$-type (c) CDW states. 
In this perspective the separation of bands into \textit{in~plane} and \textit{out~of~plane} character is even more evident, as CDW order splits the FS sheets into pockets of uniform character by removing a part of the quasi-planar sheets which connect them. 
This effect is already in action in the $q_{1c}$-type CDW, where a part of the quasi-planar sheet is already broken, and it further develops in the $q_{2c}$-type CDW state. 
Here, the FS becomes divided into a snake-like sheet connecting the pockets with \textit{in~plane} character (orange transparent box) and in the characteristic hourglass-shaped pillars with only \textit{out~of~plane} character (azure transparent box), while an isolated pocket around the Z-point remains essentially unchanged.

\begin{figure}[ht]
\includegraphics[angle=0,width=1.0\columnwidth]{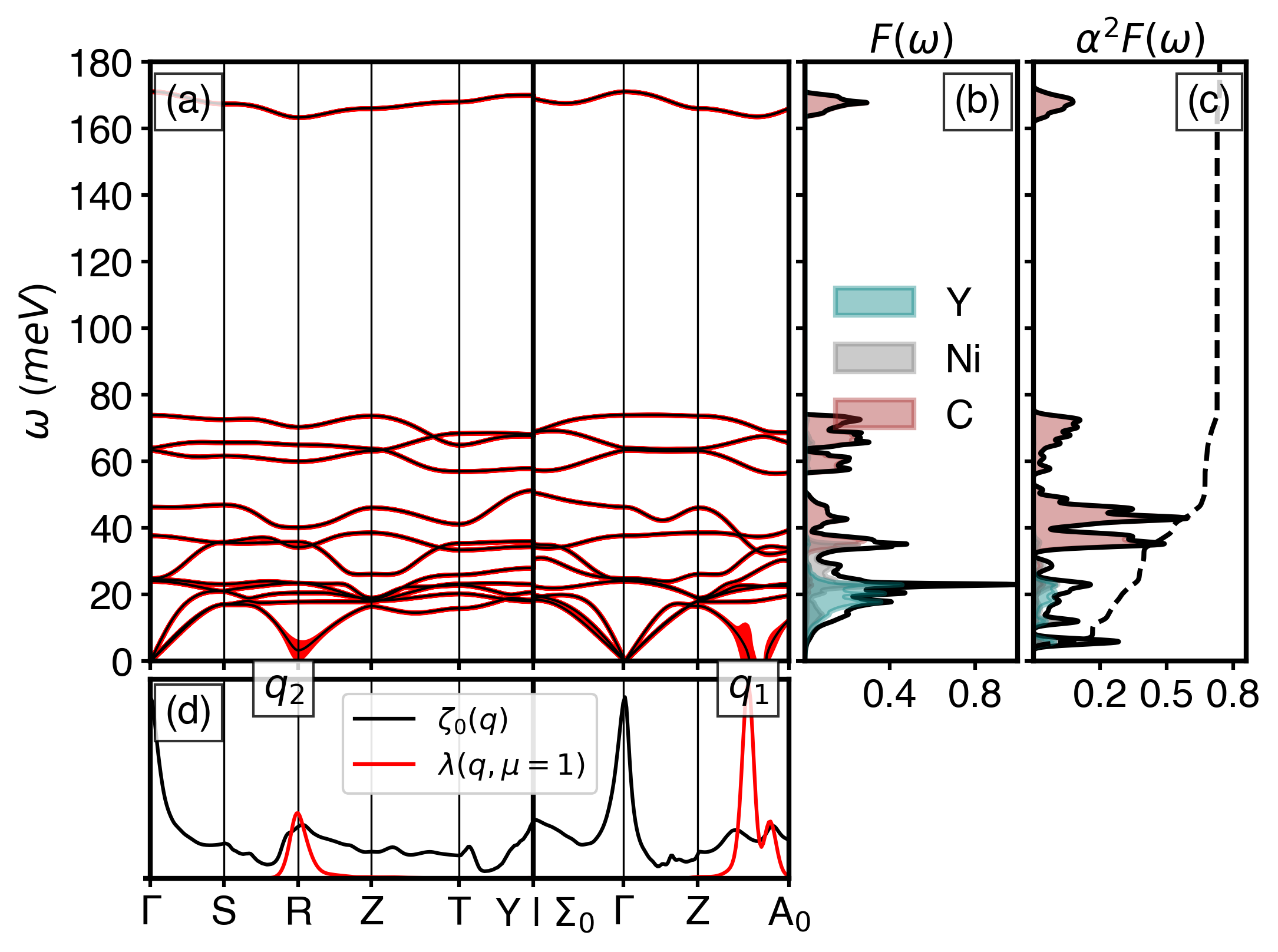}
 \caption{\label{fig:phonons_ortho} Phonon dispersions (a), decorated with electron-phonon coupling (red thick line), atom-projected phonon DOS, $F(\omega)$ (b), and electron-phonon spectral function, $\alpha^2F(\omega)$ (c), with integrated electron-phonon coupling. 
 The projections onto Y, Ni, and C, are indicated as turquise, gray, and red shaded areas. 
 (d) static limit of the bare susceptibility, $\zeta_{0}$, black line, and electron-phonon coupling for the first phonon mode, $\lambda(\mu=1)$, red line.}
  \end{figure}

Having examined the direct effect of the CDWs onto the electronic properties, we now move to establish its connection with the lattice dynamics. 
Figure~\ref{fig:phonons_ortho} displays the phonon dispersions (a) calculated for the orthorhombic parent structure, along with the atom-projected phonon DOS (b) and the electron-phonon spectral function (c) and the nesting function $\zeta_{0}$ ({\emph i.e.}\ the static limit of the bare susceptibility) (d), calculated as defined in Ref.~\cite{ponce_epw_2016}.  
The latter quantity shows two broad, hump-shaped maxima located close to the points corresponding to the $q_1$ and $q_2$ vectors. 
The broadening is particularly pronounced in the vicinity to $q_1$, where the maximum splits into two close peaks.  
The $k$-space shape of $\zeta_{0}(q)$ has previously been suggested to determine the presence or absence of CDW in early-lanthanide based $R$NiC$_2$ \cite{Laverock2009}. 
While sharp and well defined $\zeta_{0}(q)$ peaks at $q_1$ were found to promote this type of CDW in NdNiC$_2$, GdNiC$_2$ and SmNiC$_2$, the broadened features, reminiscent to those reported here for YNiC$_2$, were seen in the case of LaNiC$_2$, which shows no CDW order.

The phonon dispersions displayed in Fig.~\ref{fig:phonons_ortho}a present two dips: one at the R point ($q_{2}$), and one halfway between Z and A$_{0}$ ($q_{1}$). 
As highlighted by the thick red line around the dispersion, and the red curve below, in these two points we find the largest mode-resolved electron-phonon coupling $\lambda_{q, \mu = 1}$. 
We note that the electron-phonon coupling is strongly anisotropic, and directly correlates with the orbital character (see SM~\cite{SM}, Fig.~S8 for a direct comparison between the orbital character and the electron-phonon coupling)~\cite{an2001superconductivity, heil2015influence, heil2019superconductivity}. 
The two maxima of $\lambda_{q, \mu = 1}$ ultimately dictate the CDW modulation vectors, rather than nesting alone, as argued in a general framework in Ref.~\cite{Johannes2008}.

These results allow us to confirm that the tendency to form both CDW phases is inherent to the orthorhombic parent phase. 
In addition, phonon calculations for the two CDW states (see SM~\cite{SM}, Fig.~S9) reveal that the precursors of the $q_{1c}$-type CDW are present in the $q_{2c}$-type one, and vice versa. 
Since all three phases are almost degenerate in energy (see Tab. III of SM~\cite{SM}), and the depletion region of the electronic DOS in the $q_{1c}$-type CDW state is narrower than that of the $q_{2c}$-type CDW, the tendency of the $q_{1ic}$-type CDW to form at higher temperature than the $q_{2c}$-type can be qualitatively explained as a kind of intermediate phase with a slightly higher electronic entropy gain (see Fig.~S7 of SM~\cite{SM}).

\section{Summary and discussion}

Shedding a light on the interplay between the competing electronic instabilities in $R$NiC$_2$ family has been a long-awaited step towards understanding the underlying physics. 
Such an opportunity has been opened by recent obtaining of high quality single crystals, allowing us for a direct and detailed exploration of the character and mechanisms of two CDW-types occurring in YNiC$_2$ probed by the combination of various experimental techniques with complementary theoretical calculations.
The two types of CDW phases resolved for YNiC$_2$, and characterized by wavevectors $q_{1ic}$ and $q_{2c}$, leave distinct footprint in crystal and electronic structures and resulting experimental observables. 

The $q_{1ic}$-type CDW phase, appearing as the first CDW upon cooling from the orthorhombic parent state below $T_{1ic}$, weakly modifies the crystal structure producing a small incommensurate lattice modulation. 
The onset of this state opens a nearly isotropic electronic gap at Fermi energy. 
While a part of the high temperature FS is nested and removed, its overall character and topography is mostly conserved. 
The quasi-planar FS sheets running perpendicular to a$^*$ of the undistorted orthorhombic state, responsible for favoring the $a$ direction in the terms of electrical resistivity, retain most of their area. 
For that reason, the $\rho_b$/$\rho_a$ and $\rho_c$/$\rho_a$ ratios, being the effective measure of transport anisotropy, remain rigid across the transition towards the $q_{1ic}$-type CDW state.

The $q_{2c}$-type CDW, which prevails over the $q_{1ic}$-type CDW phase at lower temperatures much more significantly modifies the properties of YNiC$_2$. 
The crossover between the two CDW types leads to a large modification of the underlying crystal lattice, to the extent that it lowers the symmetry class from \textit{Amm2} to \textit{Cm}. 
Moreover, the $q_{2c}$-type CDW state causes a more drastic FS decomposition. 
This process removes a large part of quasi-planar sheets, which in both, the orthorhombic undistorted and the $q_{1c}$-approximate CDW state, bridge the subsequent FS counterparts. 
As a consequence, FS in the $q_{2c}$-type CDW state is reduced to smaller, yet multiple, isolated pockets, predominantly with 3D character and dispersion in all three dimensions.
This phenomenon is directly reflected by the suppression of electronic anisotropy as observed in transport properties. 
In particular, one can notice, that the thermal dependence of $\rho_a$ experiences a pronounced upturn at $T_{2c}$, which is caused by removal of large FS part perpendicular to $a^*$. 
Simultaneously, $\rho_b(T)$ and $\rho_c(T)$ turn in the opposite direction, which is caused by the increased contribution of quasi-3D FS elements.
Moreover, the contrast in the global influence on the electronic structure is demonstrated by the larger, as compared to the $q_{1c}$-approximate CDW state, decrease of DOS in the vicinity of the Fermi level, as shown in Fig.~S6 of SM~\cite{SM}. 

Although the impact on the FS differs significantly between the two CDW phases, there exists a common FS part, which is decomposed in each of them. 
This is further supported by the bare susceptibility $\zeta_0$, which was calculated within both states, $q_{1c}$- and $q_{2c}$-type (see Fig.~S9 in SM~\cite{SM}), and shows no clear peaks corresponding to the complementary CDW phase. 
The opening of a gap significantly reduces the potential for a secondary FS modification caused by nesting of that part which survives the first CDW transition.
Therefore, YNiC$_2$ does not belong to the group of systems where the onset of a secondary density wave instability complements the FS decomposition already inflicted by the preexisting CDW, and removes its remaining components. 
Such an avalanche of two or three subsequent CDW instabilities has been reported in tungsten bronzes, where, in contrast to the case of YNiC$_2$, the independent orders supplement each other and do not compete for the same FS spot \citep{Wang_1989, Canadell_1991}.

An intriguing observation stems from the comparison of the electron-phonon coupling parameter, $\lambda$, calculated in the $q_{1c}$-approximate and the $q_{2c}$-type CDW states, respectively. 
Within each of the two CDW states, $\lambda(q)$ shows distinct maxima corresponding to the alternative CDW modulation. 
In particular, within the $q_{1c}$--approximate CDW state, $\zeta_0(q)$ shows a weakly developed feature at $q\simeq q_{2}$, while a well-defined, sharp peak is seen in $\lambda(q)$ at the same $q$-position. 
This observation suggests that momentum dependent electron-phonon coupling plays a decisive role in the first-order incommensurate to commensurate CDW phase transition occurring at $T_{2c}$. 
Moreover, also in the $q_{2c}$-type CDW state, one can notice a clear maximum of $\lambda(q)$ at $q_1$.
Interestingly, the signatures of such dormant adversary state precursors have already been observed in SmNiC$_2$ and GdNiC$_2$ compounds \citep{Shimomura2009, Shimomura2016}.
In these materials, where only the $q_{1ic}$-type CDW develops a long range order, a weak X-ray diffuse scattering signal remains observable at the $q_{2c}$-position. 
Therefore, in YNiC$_2$, and likely also in other intermediate members of the $R$NiC$_2$ series, small details may decide on the result of the competition between $q_{1ic}$ and $q_{2c}$ lattice modulations. 
The suppression of one of these entities by internal or external factors, such as rare-earth magnetism, uni-axial pressure or strain, may give the other one a potential to take over.

\section{Conclusions}
We have explored physical properties of YNiC$_2$, with particular emphasis on investigating the character, origin and consequences of the two charge density wave transitions observed in this compound. 
To obtain a detailed picture we have combined a number of different experimental techniques and theoretical methods, from single crystal diffraction, through thermal and transport studies to DFT calculations. 
Our analysis shows the contrasting effects of the two CDW states examined.
The first transition near room temperature, from the undistorted orthorhombic CeNiC$_2$-type parent structure towards the CDW state with modulation vector $q_{1ic} = (0.5, 0.5+\eta, 0)$, only moderately impacts the electronic structure and related physical properties.
The consequences of the second CDW transition to a state characterized by the modulation vector $q_{2c} = (0.5, 0.5, 0.5)$ are more pronounced.
It not only changes the crystal structure by lowering the symmetry class and by generating larger atomic displacements, but also leads to a more significant FS decomposition, which is reflected in a considerably reduced anisotropy of transport properties. 
Both types of CDW instabilities are triggered by both FS nesting and momentum dependent electron-phonon coupling. 
This common thread, despite the large contrast in the microscopic and macroscopic consequences imposed by $q_{1ic}$- and $q_{2c}$-type CDW states, suggests that the balance between electronic and lattice degrees of freedom plays an important role in the competing interplay between them.

\section{Acknowledgments}
Financial support for M.\,R.\ by Grant No.\ BPN/BEK/ 2021/1/00245/DEC/1 of the Bekker Program of the Polish National Agency for Academic Exchange (NAWA) is gratefully acknowledged. 
S.\,D.\,C.\ acknowledges funding from the European Union - NextGenerationEU under the Italian Ministry of University and Research (MUR), “Network 4 Energy Sustainable Transition - NEST” project (MIUR project code PE000021, Concession Degree No.~1561 of Oct.\,11, 2022)  CUP C93C22005230007. 
Metallography support by Snezana Stojanovic and support with scanning electron microscopy and microprobe analysis by Monika Waas and Robert Svagera are gratefully acknowledged. 
S.\,D.\,C.\ likes to thank Lilia Boeri, Andriy Smolyanyuk and Martin Bra{\ss} for the helpful discussion.
The computational results presented have been achieved using the Vienna Scientific Cluster (VSC).

\end{document}


\title{Supplemental Material for: \\
Competing charge density wave phases in YNiC$_2$}

\author{Marta Roman}
\email{marta.roman@pg.edu.pl}
\affiliation{Institute of Solid State Physics, TU Wien, Wiedner Hauptstrasse 8-10, A-1040 Wien, Austria}
\affiliation{Institute of Physics and Applied Computer Science, Faculty of Applied Physics and Mathematics, Gdansk University of Technology,
Narutowicza 11/12, 80-233 Gdansk, Poland}

\author{Simone Di Cataldo}
\affiliation{Institute of Solid State Physics, TU Wien, Wiedner Hauptstrasse 8-10, A-1040 Wien, Austria}
\affiliation{Dipartimento di Fisica, Sapienza University of Rome, Piazzale Aldo Moro 5, 00185 Rome, Italy}

\author{Berthold St{\"o}ger}
\affiliation{X-Ray Center, TU Wien, Getreidemarkt 9, A-1060 Wien, Austria}

\author{Lisa Reisinger}
\affiliation{Institute of Solid State Physics, TU Wien, Wiedner Hauptstrasse 8-10, A-1040 Wien, Austria}

\author{Emilie Morineau}
\affiliation{Institute of Solid State Physics, TU Wien, Wiedner Hauptstrasse 8-10, A-1040 Wien, Austria}

\author{Kamil K. Kolincio}
\affiliation{Faculty of Applied Physics and Mathematics, Gdansk University of Technology,
Narutowicza 11/12, 80-233 Gdansk, Poland}

\author{Herwig Michor}
\email{michor@ifp.tuwien.ac.at}
\affiliation{Institute of Solid State Physics, TU Wien, Wiedner Hauptstrasse 8-10, A-1040 Wien, Austria}

\maketitle

\section{Computational Details}
\label{suppsect:computational_details}
All calculations were performed using Quantum ESPRESSO \cite{Giannozzi_JPCM_2009_qe, Giannozzi_JPCM_2017_qe} and EPW \cite{ponce_epw_2016}. We employed Optimized Norm-Conserving Vanderbilt Pseudopotentials \cite{Hamann_PRB_2013_ONCV}, using the Perdew-Burke-Ernzerhof approximation for the exchange correlation functional. We employed a cutoff of 80 Ry on the plane waves expansion. 

\begin{itemize}
 \item \textbf{Orthorhombic parent phase}, self-consistency was achieved employing a uniform 12$\times$12$\times$12 Monkhorst-Pack mesh and a Methfessel-Paxton smearing of 0.02 Ry for Brillouin zone integration of the ground-state charge density. The density of states was calculated non-self-consistently over a 32$\times$32$\times$32 grid, using the tetrahedron method for integration. Phonon properties were computed over a 4$\times$4$\times$4 grid. Using EPW, they were then interpolated over a 32$\times$32$\times$32 $k$ grid for the calculation of the phonon selfenergy and the bare susceptibility. Wannierization was obtained from a non-self-consistent calculation over a 4$\times$4$\times$4 $k$-grid, using an initial guess with 5 $d$ orbitals centered on Ni, and three $sp^{2}$ orbitals centered in between the two carbon atoms. An outer window from -4 to 4 eV, and a frozen window from -0.8 to 1.3 eV were employed for disentanglement.
  \item \textbf{q$_{1c}$-type CDW phase}, self-consistency was achieved employing a uniform 12$\times$12$\times$12 Monkhorst-Pack mesh and a Methfessel-Paxton smearing of 0.02 Ry for Brillouin zone integration of the ground-state charge density. The density of states was calculated non-self-consistently over a 24$\times$24$\times$24 grid, using the tetrahedron method for integration. 
  \item \textbf{q$_{2c}$-type CDW phase}, self-consistency was achieved employing a uniform 4$\times$4$\times$4 Monkhorst-Pack mesh and a Methfessel-Paxton smearing of 0.04 Ry for Brillouin zone integration of the ground-state charge density. The density of states was calculated non-self-consistently over a 24$\times$24$\times$24 grid, using the tetrahedron method for integration. 
 \end{itemize}

\section{Details on the lattice notation and CDW wavevectors}
\label{sect:details_notation}
In this section we clarify the settings employed in our calculations to describe coordinates and crystal structures. 
\subsection{Transformation from $Amm2$ to $Cm2m$}

In the literature, $R$NiC$_2$ in the orthorhombic phase are often described in terms of their \textit{conventional} cell in the $Amm2$ setting ($a < b < c$), i.e. with $a = 3.57$ \AA, $b = 4.51$ \AA, and $c = 6.03$ \AA. \textbf{Note:} in the rest of this section and throughout the paper, $a$, $b$, and $c$ remain fixed to these values.

Despite the $Amm2$ setting is often adopted in the experimental literature, we chose to employ the $Cm2m$ setting, following the notation of Cracknell {\sl et al.}~\cite{Cracknell_Book_IrrepsTables_1979} and employed in the Bilbao Crystallographic Server \cite{Aroyo_Bilbao1_2006, Aroyo_Bilbao2_2006}.

In this setting, the lattice vectors take the form:
\begin{gather}
\label{eq:matrix_amm2}
\mathcal{A}^{p}_{Amm2} = 
\begin{bmatrix} 
\vec{a}_1 \\
\vec{a}_2 \\
\vec{a}_3 \\
\end{bmatrix}
 =
  \begin{pmatrix}
 a/2  &  -b/2  &  0 \\
 a/2  &   b/2  &  0 \\
   0  &   0    &  c \\
   \end{pmatrix}
\end{gather}

\begin{gather}
\label{eq:matrix_cm2m}
\mathcal{A}^{p}_{Cm2m} = 
\begin{bmatrix} 
\vec{a}_1 \\
\vec{a}_2 \\
\vec{a}_3 \\
\end{bmatrix}
 =
  \begin{pmatrix}
 b/2  &  -c/2  &  0 \\
 b/2  &   c/2  &  0 \\
   0  &   0    &  a \\
   \end{pmatrix}
\end{gather}

The corresponding reciprocal lattice vectors are given by ($\mathcal{A}$ and $\mathcal{B}$ denote matrices for direct and reciprocal lattices, respectively, the apex $^{p}$ denotes that these refer to the \textit{primitive} lattice vectors):
\begin{gather}
\label{eq:matrix_recip_amm2}
\mathcal{B}^{p}_{Amm2} = 
\begin{bmatrix} 
\vec{b}_1 \\
\vec{b}_2 \\
\vec{b}_3 \\
\end{bmatrix}
 = 2\pi
  \begin{pmatrix}
 1/a  &   1/a  &  0   \\
-1/b  &   1/b  &  0   \\
   0  &   0    &  1/c \\
   \end{pmatrix}
\end{gather}

\begin{gather}
\label{eq:matrix_recip_cm2m}
\mathcal{B}^{p}_{Cm2m} = 
\begin{bmatrix} 
\vec{b}_1 \\
\vec{b}_2 \\
\vec{b}_3 \\
\end{bmatrix}
 = 2\pi
  \begin{pmatrix}
 1/b  &   1/b  &  0   \\
-1/c  &   1/c  &  0   \\
   0  &   0    &  1/a \\
   \end{pmatrix}
\end{gather}

Given the matrix:
\begin{gather}
\mathcal{R}
 = 
  \begin{pmatrix}
 0  &  1  &  0   \\
 0  &  0  &  1   \\
 1  &  0  &  0 \\
   \end{pmatrix}
\end{gather}

and the transform matrix $\mathcal{T}$ from conventional $\mathcal{B}^{c}_{Amm2}$ to primitive $\mathcal{B}^{p}_{Amm2}$:
\begin{gather}
\mathcal{T}
 = 
  \begin{pmatrix}
 1  &  -1  &  0   \\
 1  &   1  &  0   \\
 0  &   0  &  1 \\
   \end{pmatrix}
\end{gather}
one can write the corresponding wavevectors in the primitive $Cm2m$ setting as:
\begin{equation}
\label{eq:reciprocal_lattice_conversion}
\mathcal{B}^{p}_{Cm2m} = \mathcal{T} \cdot \mathcal{R} \cdot \mathcal{B}^{c}_{Amm2} \cdot \mathcal{R}^{-1}
\end{equation}

The usual definition of $q_{1c} = \left(0.50, 0.50, 0.00\right)$ and $q_{2c} = \left(0.50, 0.50, 0.50\right)$ refers to the \textit{conventional} cell in $Amm2$ setting. As atomic coordinates are contravariant with respect to a change of basis, these wavevectors in the primitive $Cm2m$ cell are:
\begin{equation}
\label{eq:q_vectors_conversion}
\vec{q}^{p}_{Cm2m} = \mathcal{R} \cdot \vec{q}^{c}_{Amm2} \cdot \mathcal{R}^{-1} \cdot \mathcal{T}^{-1} 
\end{equation}
From which we obtain the relations in the primitive, $Cm2m$ cell.
\begin{equation}
\label{eq:q1_q2_cmm2}
\begin{split}
\vec{q}_{1c} = \left( 0.25, 0.25, 0.50 \right)  \\
\vec{q}_{2c} = \left( 0.50, 0.00, 0.50 \right) 
\end{split}
\end{equation}
\subsection{Matching $q_{1c}$/$q_{2c}$ and orthorhombic axes}
In this section we report the lattice vectors and the coordinates of the special points for the $q_{1c}$ and $q_{2c}$ phases, as well as a technical discussion on how to employ them. The coordinates employed are summarized in Tab. \ref{supptab:q1cq2c_coords}.

These coordinates can be obtained from the transform matrix $\mathcal{T}$ between the orthorhombic and the $q_{1c}$ and $q_{2c}$ lattice vectors, following precisely the same procedure described in Sect. \ref{sect:details_notation}. However, there is a technical difficulty that has to be taken care of. Density Functional Theory codes, as well as programs for structure visualization (e.g.\ VESTA \cite{momma2008vesta} will define lattice vectors following a certain convention, which effectively imposes a rotation of the Cartesian frame of reference in which the crystal is defined. In order for the procedure described in Sect. \ref{sect:details_notation} to work, one has to first establish a common frame of reference between the two crystals, and explicitly enforce the choice of lattice vectors so that the "real space" crystal described by them is oriented in the same way. Since, to the best of our knowledge, it is not possible to enforce this {\sl a priori}, we developed a python script to rotate the frame of reference of the $q_{1c}$/$q_{2c}$ crystal axes until all the atoms matched the parent phase within a certain distance threshold. 

The crystal axes and atomic coordinates for the $q_{1c}$ and $q_{2c}$ phases employed in this work are 

\begin{gather}
\label{eq:matrix_amm2}
\mathcal{A}_{q1c} = 
\begin{bmatrix} 
\vec{a}_1 \\
\vec{a}_2 \\
\vec{a}_3 \\
\end{bmatrix}
 =
  \begin{pmatrix}
 -4.524  &  0.000  &  -3.548 \\
  4.489  &  0.035  &  -3.592 \\
  0.023  & -6.027  &  -0.030 \\
   \end{pmatrix}
\end{gather}

\begin{table}[h]
    \centering
    \begin{tabular}{|c|c|c|c|}
\hline
Atom & x & y & z \\
\hline
Y   &  0.000     &    0.000     &    0.000 \\
Y   &  0.246     &    0.754     &    0.500 \\
Y   &  0.496     &    0.504     &    0.003 \\
Y   &  0.750     &    0.250     &    0.503 \\
Ni  &   0.253    &    0.257     &    0.613 \\
Ni  &   0.993    &    0.497     &    0.113 \\
Ni  &   0.503    &    0.007     &    0.113 \\
Ni  &   0.743    &    0.747     &    0.613 \\
C   &  0.071     &    0.425     &    0.802 \\
C   &  0.175     &    0.330     &    0.302 \\
C   &  0.670     &    0.825     &    0.302 \\
C   &  0.575     &    0.929     &    0.802 \\
C   &  0.824     &    0.678     &    0.302 \\
C   &  0.421     &    0.076     &    0.802 \\
C   &  0.924     &    0.578     &    0.802 \\
C   &  0.322     &    0.176     &    0.302 \\
\hline
    \end{tabular}
    \label{tab:q1cpos}
\end{table}

\begin{gather}
\label{eq:matrix_amm2}
\mathcal{A}_{q2c} = 
\begin{bmatrix} 
\vec{a}_1 \\
\vec{a}_2 \\
\vec{a}_3 \\
\end{bmatrix}
 =
  \begin{pmatrix}
  3.763  &  0.000  &   0.000 \\
 -1.068  &  3.606  &  -3.569 \\
 -1.068  &  3.606  &   3.569 \\
   \end{pmatrix}
\end{gather}

\begin{table}[h]
    \centering
    \begin{tabular}{|c|c|c|c|}
\hline
Atom & x & y & z \\
\hline
Y   &  0.005     &    0.505     &    0.505 \\
Y   &  0.995     &    0.995     &    0.995 \\
Ni  &  0.390    &     0.066    &     0.547 \\
Ni  &  0.390     &    0.547     &    0.066 \\
C   &  0.549     &    0.830     &    0.322 \\
C   &  0.549     &    0.322     &    0.830 \\
C   &  0.851      &   0.977      &   0.474 \\
C   &  0.851     &    0.474     &    0.977 \\
\hline
    \end{tabular}
    \caption{Atomic positions for $q_{2c}$ in crystal coordinates}
    \label{tab:q2cpos}
\end{table}

\begin{table}[ht]
    \centering
    \begin{tabular}{|c|c||c|c|}
\hline
$q_{1c}$   &                     & $q_{2c}$   &                      \\
\hline
\hline
 Label     &  Coordinates         & Label     &  Coordinates\\
 \hline
$\Gamma$     &(0.00,0.00,0.00)    &$\Gamma$     &(0.00,0.00,0.00)     \\
S            &(-0.50,0.50,-0.50)    &S            &(-0.05,0.51,0.51)     \\
R            &(-1.00,0.00,-0.50)    &R            &(-0.05,0.01,1.01)     \\
Z            &(-0.50,-0.50,0.00)    &Z            &(0.00,-0.50,0.50)     \\
T            &(-0.50,-0.50,-1.00)    &T            &(-0.53,-0.04,0.96)     \\
Y            &(0.00,0.00,-1.00)    &Y            &(-0.53,0.46,0.46)     \\
$\Sigma_{0}$ &(-0.78,0.78,0.00)    &$\Sigma_{0}$ &( 0.34,0.43,0.43)     \\
$\Gamma$     &(0.00,0.00,0.00)    &$\Gamma$     &( 0.00,0.00,0.00)     \\
Z            &(-0.50,-0.50,0.00)    &Z            &( 0.00,-0.50,0.50)     \\
A$_{0}$      &(-1.28,0.27,0.00)    &A$_{0}$      &( 0.34,-0.07,0.93)     \\
\hline
    \end{tabular}
    \caption{Coordinates of the special points for the $q_{1c}$ and $q_{2c}$ phases employed in the band structure and phonon calculations}
    \label{supptab:q1cq2c_coords}
\end{table}

\renewcommand{\thefigure}{S1}
\begin{figure}[ht]
\includegraphics[angle=0,width=0.98\columnwidth]{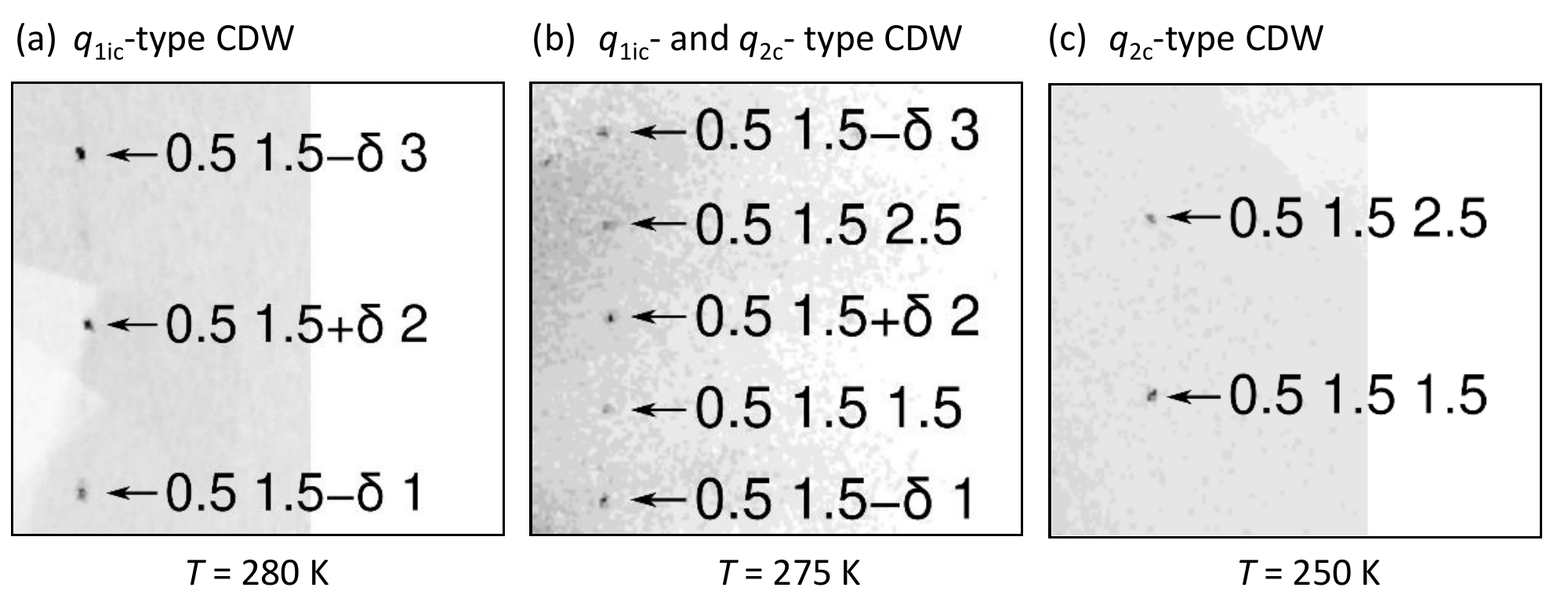}
 \caption{\label{suppfig:xrd} $k=1.5$ reciprocal space sections reconstructed from intensity data of a crystal in (a) the $q_{1ic}$-type CDW state, (b) both, the $q_{1ic}$-type and the $q_2$-type CDW states, and (c) only the $q_2$-type CDW state. Reflection indices are given with respect to the orthorhombic non-modulated structure.}
\end{figure}

\FloatBarrier
\newpage
\section{Complementary experimental results}
\label{sect:experimental results}

\subsection{Details of CDW phase transitions}

In order to probe for thermal hysteresis across the two CDW phase transitions as accurate as possible, we measured the electrical resistivity with a YNiC$_2$ single crystal bar cut along the orthorhombic $c$-axis (see results in Fig.\ref{suppfig:hist_res}). The crystal was mounted on the PPMS rotator-insert with a metal plate sample holder where the sample is glued with GE-varnish. 
For electrical insulation, a thin cigarette paper is placed in between sample and the metal plate which is directly touching the calibrated CERNOX temperature sensor of the PPMS rotator-insert, thus insuring closest thermal contact between the sample and the temperature sensor.   
For a reliable thermalisation, the temperature was fully stabilized for each measurement point, with an average speed of temperature variation of 0.15\,K/min in the $T$-range from 280 to 320 K and below 264 K, but as slow as 41\,mK/min within the  $T$-interval from 264 to 280\, K. At the second-order phase transition neat room temperature warming and cooling data match within the noise error of the measurement, whereas from data right at the first-order transition at around 272\,K a very small but nonetheless finite hysteresis of the order of 0.1\,K is revealed (see magnifying inserts in Fig.\ref{suppfig:hist_res}). 

\renewcommand{\thefigure}{S2}
\begin{figure}[ht]
\includegraphics[angle=0,width=0.8\columnwidth]{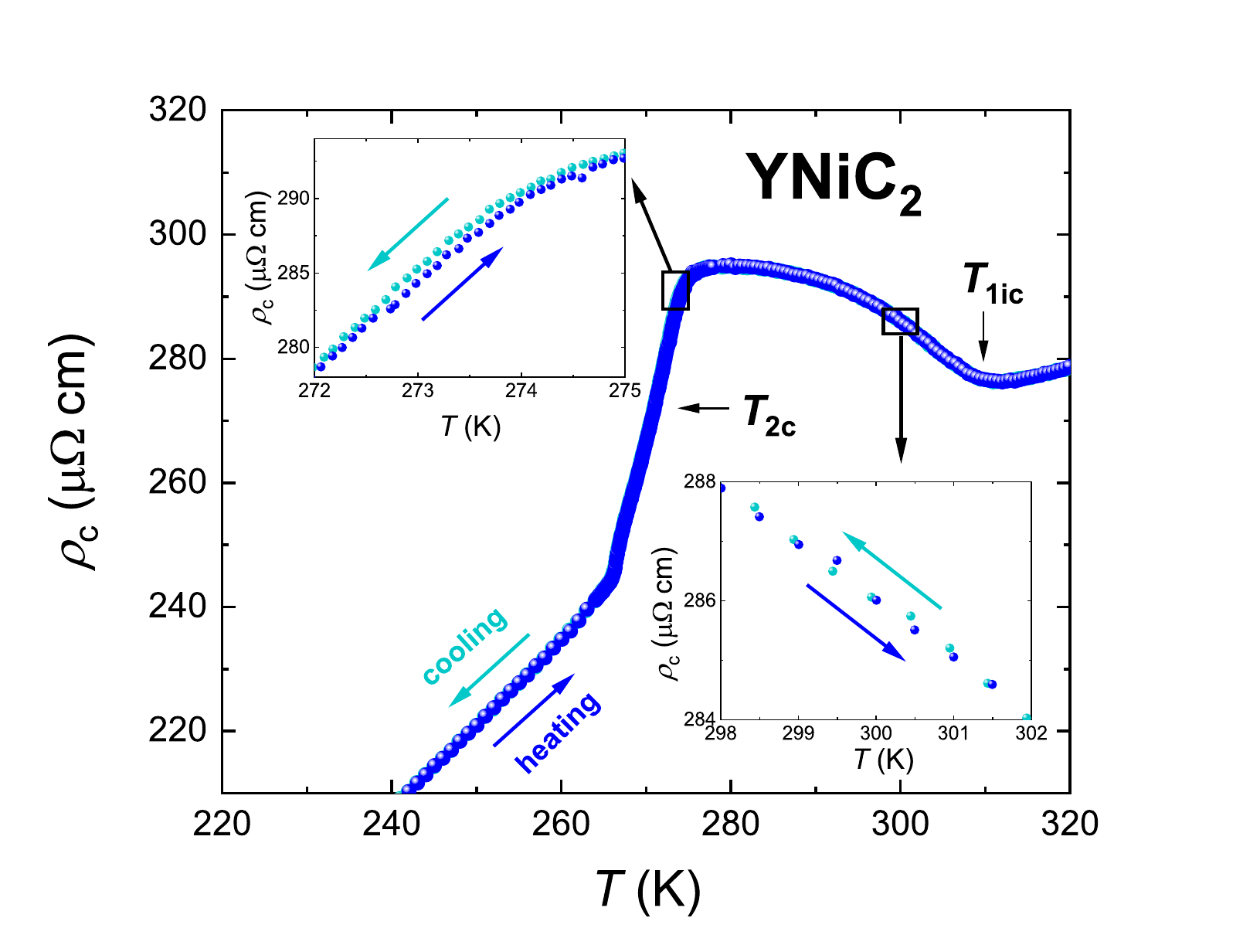}
 \caption{\label{suppfig:hist_res} High-temperature-resolution measurement of YNiC$_2$ electrical resistivity data, $\rho_c(T)$, measured across the second-order and first-order CDW phase transitions at $T_{1ic}$ and $T_{2c}$, respectively, with initial cooling and subsequent heating.}
\end{figure}

\renewcommand{\thefigure}{S3}
\begin{figure}[ht]
\includegraphics[angle=0,width=0.5\columnwidth]{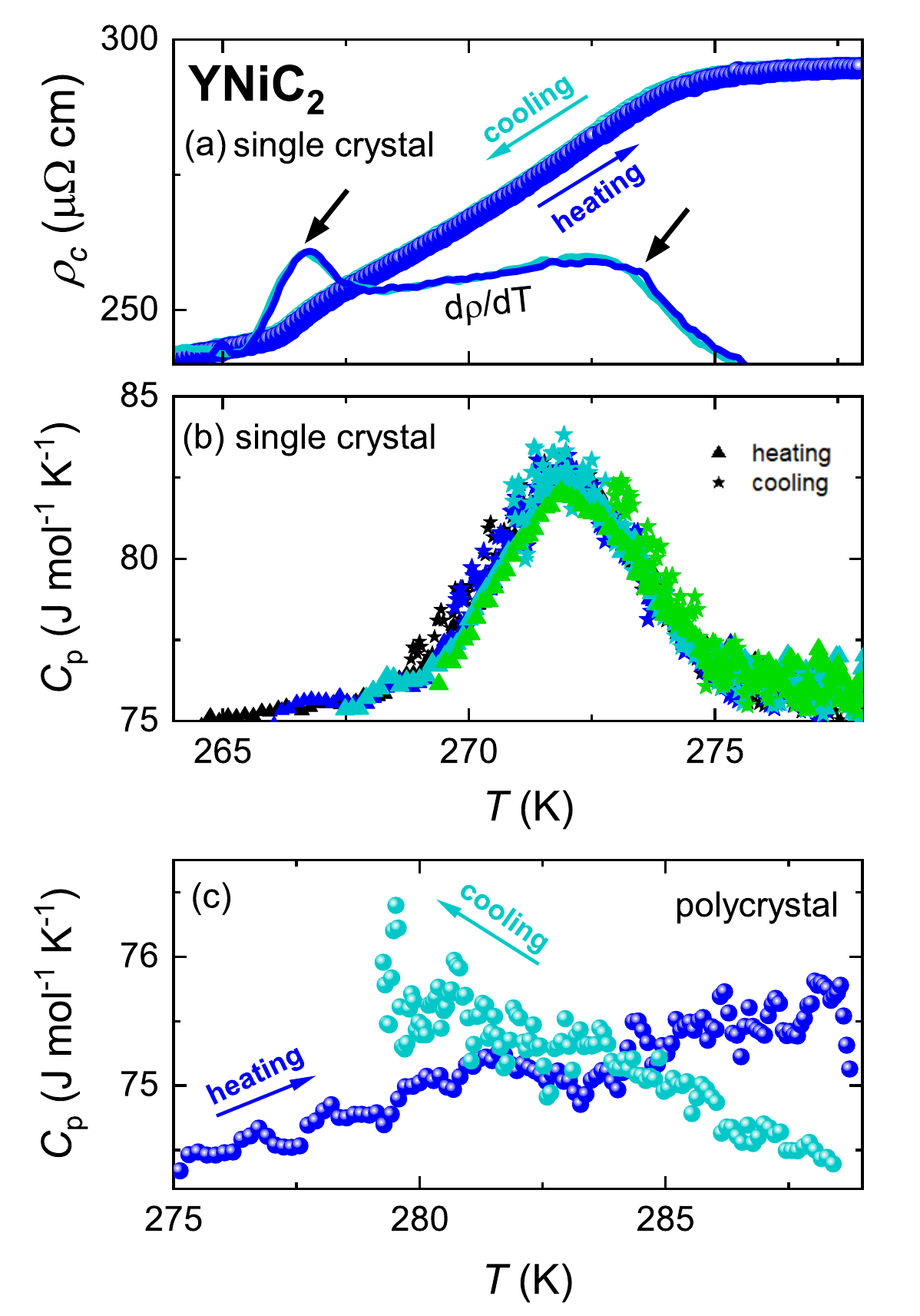}
 \caption{\label{suppfig:fot} Vicinity of the first order CDW phase transition at $T_{2c}$ in: a) electrical resistivity measured c-direction b) Specific heat anomaly related to the first order CDW transition at $T_{2c}$ as determined via a slope analysis of heating (triangle) and subsequent cooling (star) curves where different symbol colors refer to different starting temperatures of the pulses. 
 c) Specific heat anomaly of polycrystalline YNiC$_2$ related to the (further broadened) first order CDW transition at around $T_{2c}$ as determined via a slope analysis of heating (dark blue) and subsequent cooling (light blue) curves.}
\end{figure}

\renewcommand{\thefigure}{S4}
\begin{figure}[h]
\includegraphics[angle=0,width=0.5\columnwidth]{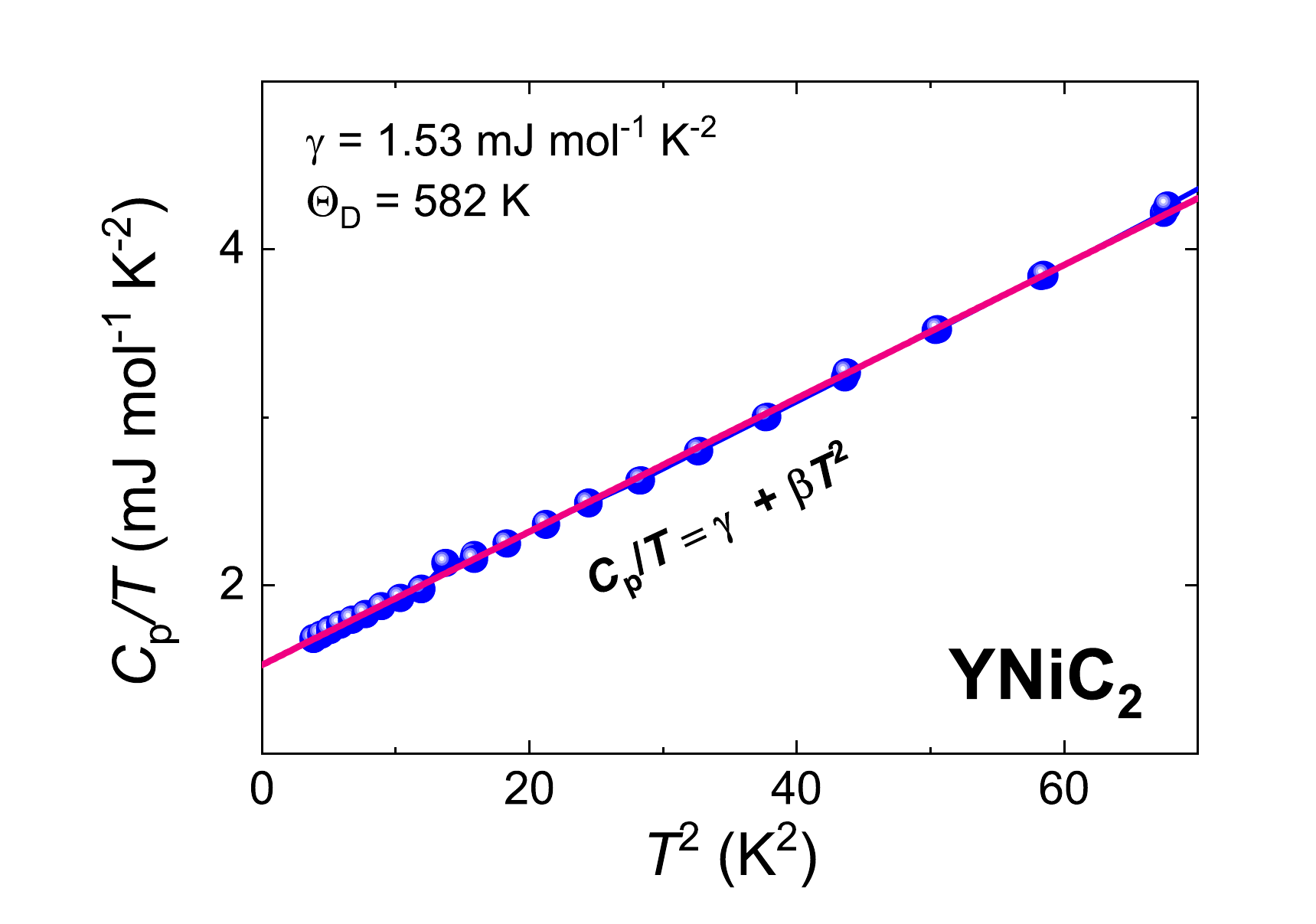}
 \caption{\label{suppfig:HClowT} C$_p/T$ vs. T$^2$ where the pink solid line represents a C$_p$/T = \(\gamma\)+{\(\beta\)}T$^2$ fit.}
\end{figure}

\subsection{Analysis of low temperature specific heat data}

The specific heat of YNiC$_2$ at temepratures below 10 K is shown as $C/T$ vs. $T^2$ plot in Fig. \ref{suppfig:HClowT} and analyzed using the formula,
\begin{equation}
\label{Cpeq}
\frac{C_p}{T} = \gamma+\beta T^2,
\end{equation}
where the first and second terms in the right side of eq. \ref{Cpeq} represent the electronic and lattice contributions according to the basic Sommerfeld and Debye models, respectively, yielding a Sommerfeld coefﬁcient $\gamma$ = 1.53  mJ mol$^{-1}$ K$^{-2}$ and Debye coefficient $\beta$ = 0.03936 mJ mol$^{-1}$ K$^{-4}$. 
The latter corresponds to a Debye temperature $\theta_{\rm D} = 582$ K according to $\theta_{D} = {\left(\frac{12 \pi^4}{5\beta}nR\right)}^{\frac{1}{3}}$ where R = 8.314 mol$^{-1}$ K$^{-1}$ and $n$ is the number of atoms per formula unit (here $n = 4$ for YNiC$_2$). 

The estimated value of Sommerfeld coefficient matches well with the value $\gamma$ = 1.67 mJ mol$^{-1}$ K$^{-2}$ determined for a polycrystalline representative \citep{Kolincio2019}. 
However, it is slightly larger than $\gamma$ = 0.83  mJ mol$^{-1}$ K$^{-2}$ for single-crystalline LuNiC$_2$, also hosting $q_{2c}$-type CDW  \citep{Steiner2018} suggesting that the electronic structure of YNiC$_2$ is possibly less affected by the $q_{2c}$-type CDW formation.

\FloatBarrier
\section{Details of electronic properties}



\renewcommand{\thefigure}{S5}
\begin{figure}[h]
\includegraphics[angle=0,width=0.65\columnwidth]{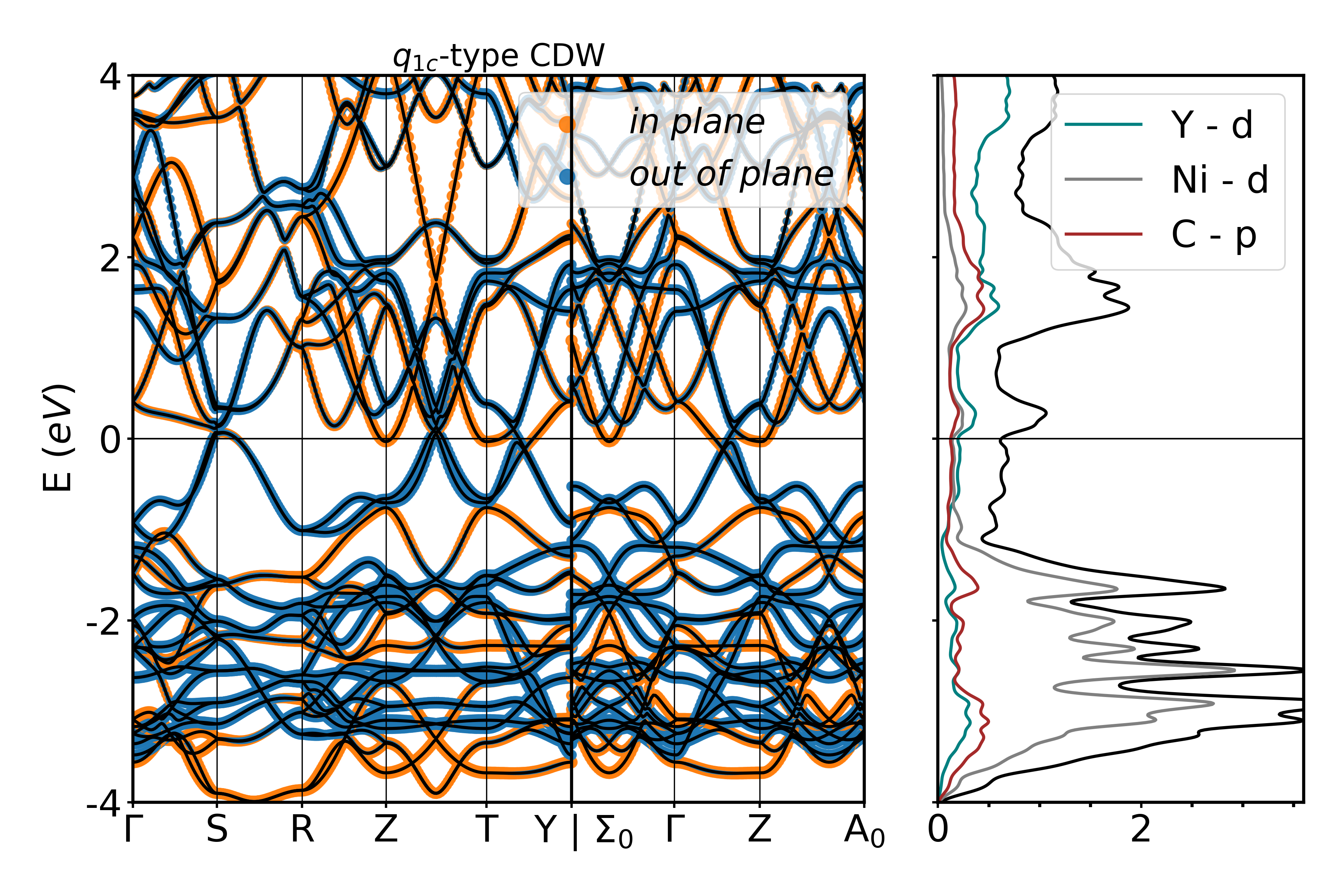}
\includegraphics[angle=0,width=0.65\columnwidth]{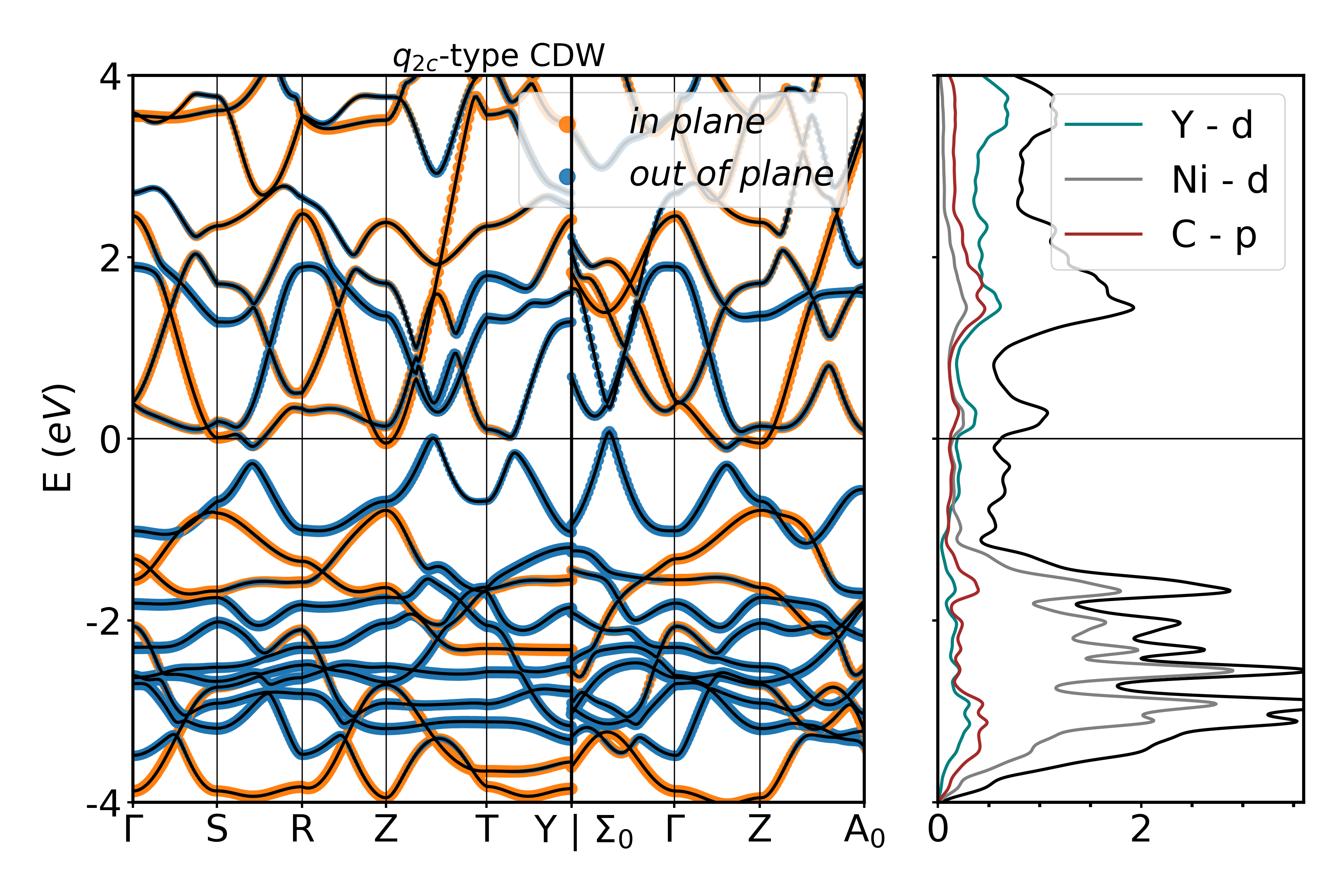}
 \caption{Electronic band structure for the $q_{1c}$-type (upper) and $q_{2c}$-type (lower) CDW phases. The bands are colored with the \textit{in plane} and \textit{out of plane} projection of orbitals, as defined in the main text.}
\end{figure}

\label{sect:details_electronic}
\begin{table}[h]
 \caption{Relative DFT total energy between the CDW and parent phase. These energy differences are close to the accuracy limit of our DFT calculations ($\sim 1$ meV/atom).}
    \centering
    \begin{tabular}{|c|c|}
    \hline
    Phase                    &  $\Delta E$ (meV/f.u.) \\
    \hline
    Ortho (parent phase)     &       0.0  \\
    q$_{1c}$-type CDW         &       2.8  \\
    q$_{2c}$-type CDW         &       6.0  \\
    \hline
    \end{tabular}
    \label{tab:energies}
\end{table}

\renewcommand{\thefigure}{S6}
\begin{figure}
\includegraphics[angle=0,width=0.35\columnwidth]{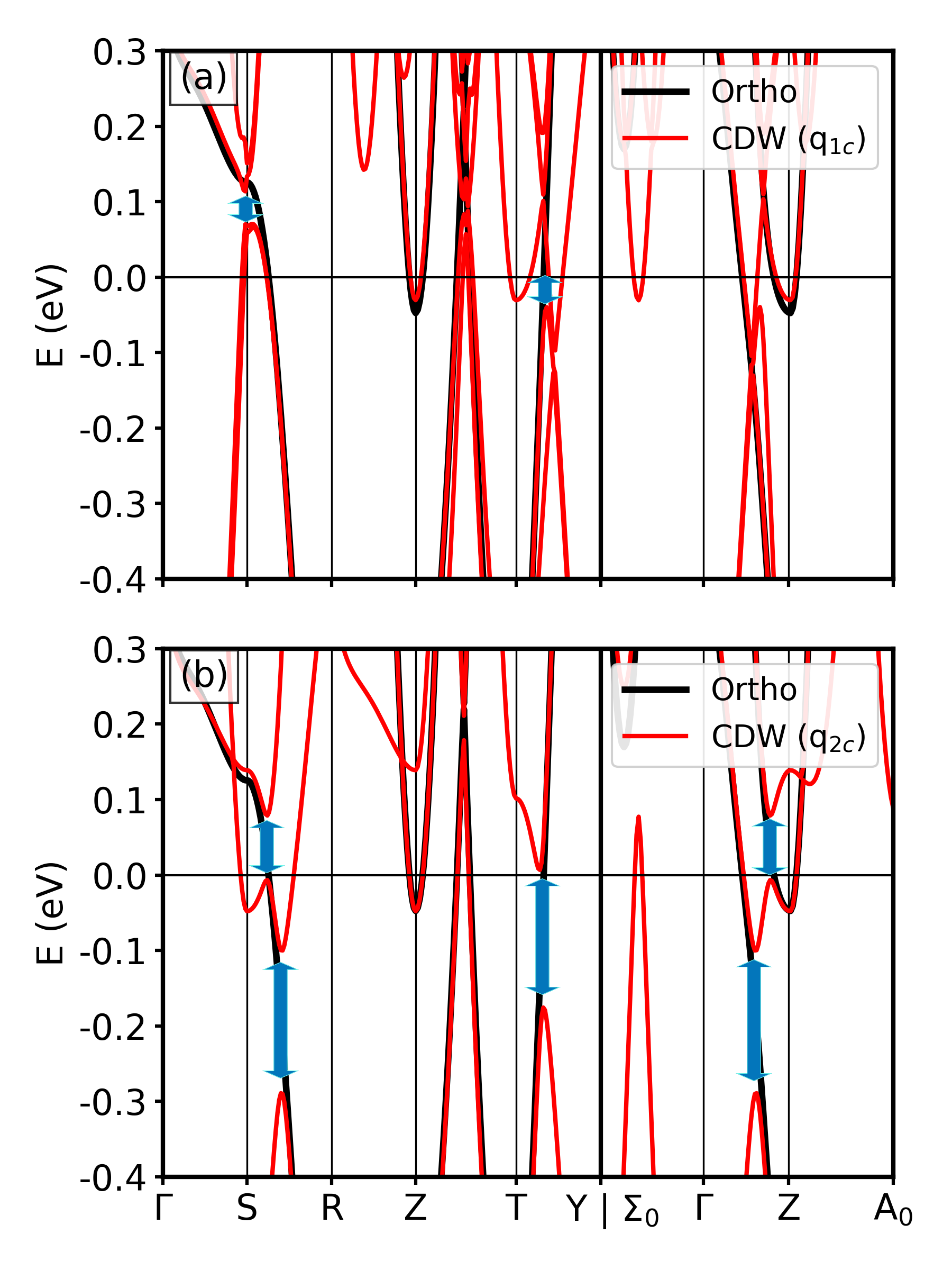}
\includegraphics[angle=0,width=0.42\columnwidth]{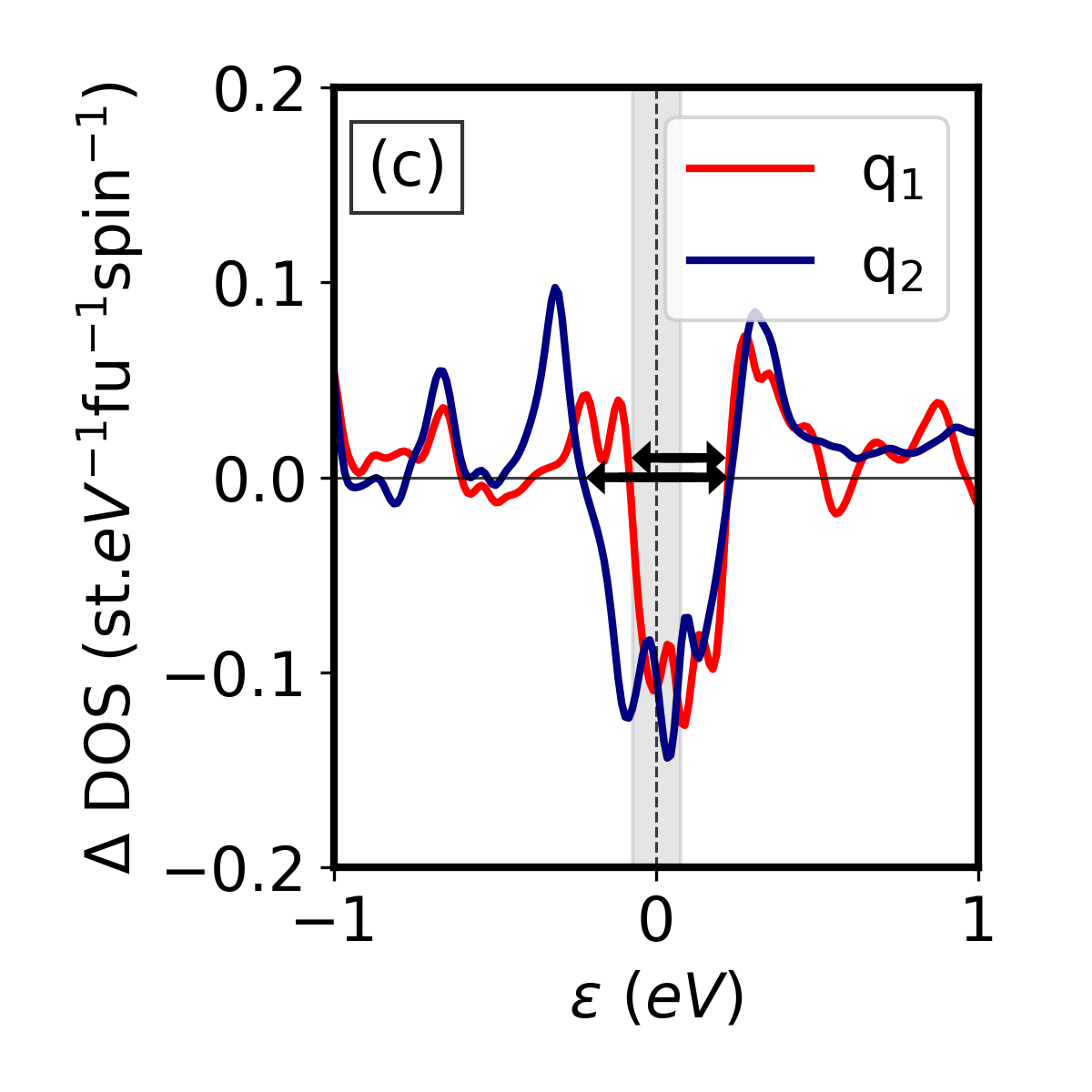}
 \caption{\label{suppfig:delta_dos} Zoom-in of the electronic band structure near the Fermi energy for the q$_{1c}$-type CDW (a) and the q$_{2c}$-type CDW (b) as compared to the orthorhombic parent phase. Blue arrows highlight CDW band splittings. (c): Difference between the DOS of the $q_{1c}$- (in red) and $q_{2c}$-type CDW phases (in blue) with respect to the orthorhombic parent phase. Horizontal black arrows highlights the CDW pseudogap (283 and 414 meV, respectively), and a vertical gray shaded area highlights the energy region of $\pm 3 k_BT$ around the Fermi energy at 300 K. }
\end{figure}

\renewcommand{\thefigure}{S7}
\begin{figure}
\includegraphics[angle=0,width=0.35\columnwidth]{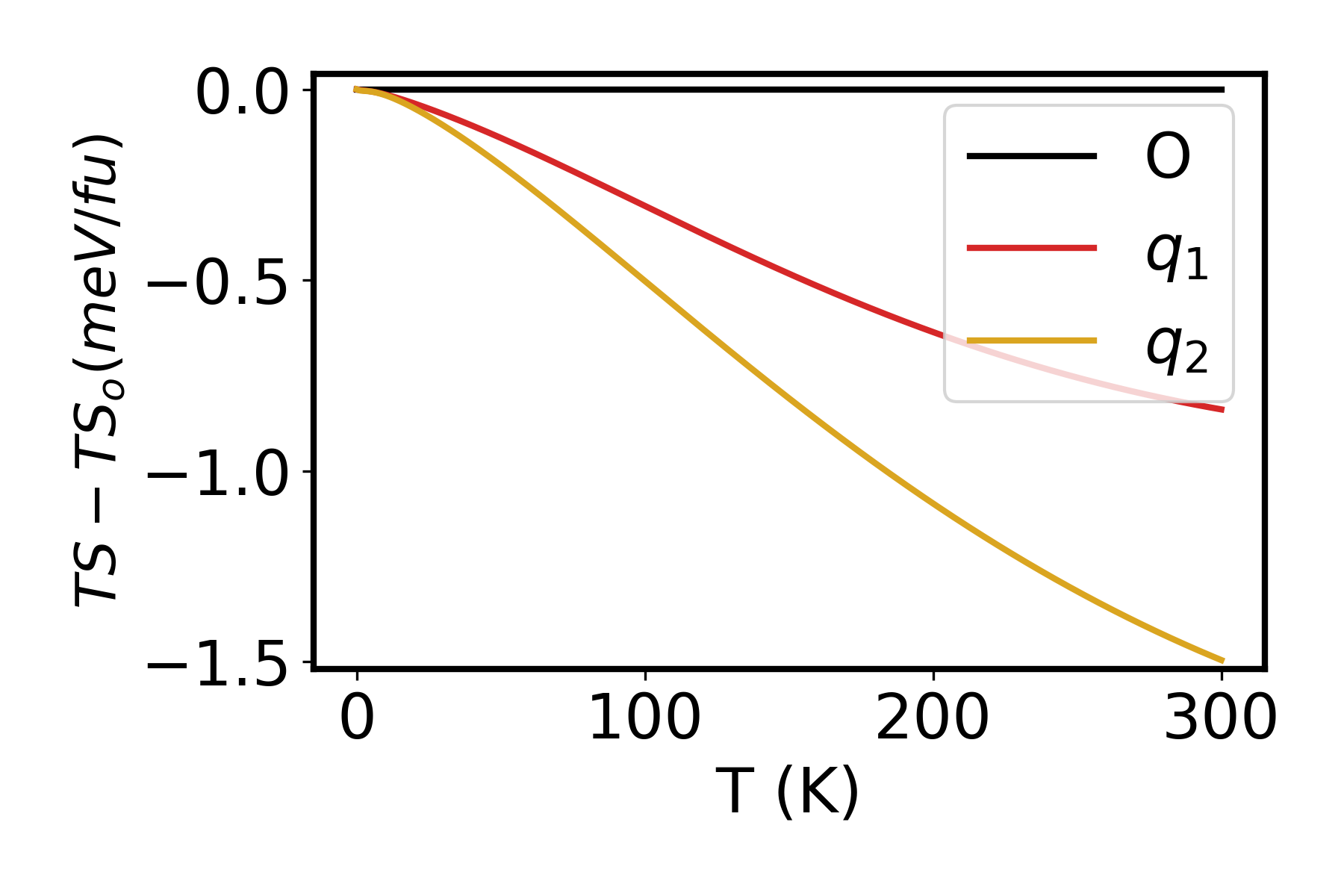}
 \caption{\label{suppfig:entropy} Electronic contribution to the entropy as a function of temperature, relative to the orthorhombic phase for the $q_{1c}$-type and $q_{2c}$-type CDW phases.}
\end{figure}

\renewcommand{\thefigure}{S8}
\begin{figure}[ht]
\includegraphics[angle=0,width=0.46\columnwidth]{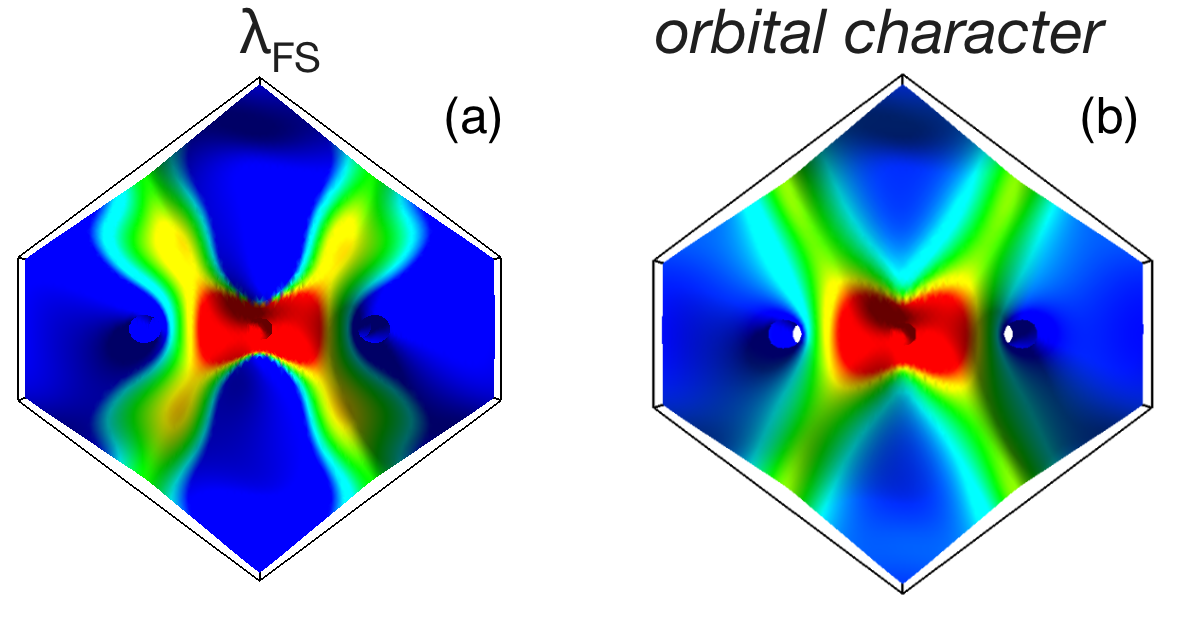}
 \caption{\label{suppfig:FS_lamda_proj_comparison} Top view comparison of the Fermi surface of the orthorhombic parent phase of YNiC$_{2}$ decorated with the anisotropic electron-phonon coupling $\lambda_{q}$ (a) and with the projection onto \textit{in plane} states (see main text for the definition) (b). The range of the color scale in (a) from blue to red corresponds to values between 0.28 and 0.35 for the coupling, and a range from \textit{out of plane} to \textit{in-plane} for the orbital projection in (b).}
\end{figure}

\FloatBarrier
\section{Details of phonon properties}
In this section we show the phonon dispersions of the $q_{1c}$-type (left) and $q_{2c}$-type (right) CDW phases, calculated from the phonon selfenergy within EPW \cite{ponce_epw_2016, lee_2023_epw}. In the bottom panel, we report the static limit of the bare susceptibility along the same path. We note that in the $q_{1c}$ phase, the instability relative to the $q_{2c}$ phase is visible, and vice versa. Note that in $q_{1c}$ the $\Gamma$ point of the parent cell is folded between $Z$ and $A_{0}$.

\renewcommand{\thefigure}{S9}
\begin{figure}[ht]
\includegraphics[angle=0,width=0.52\columnwidth]{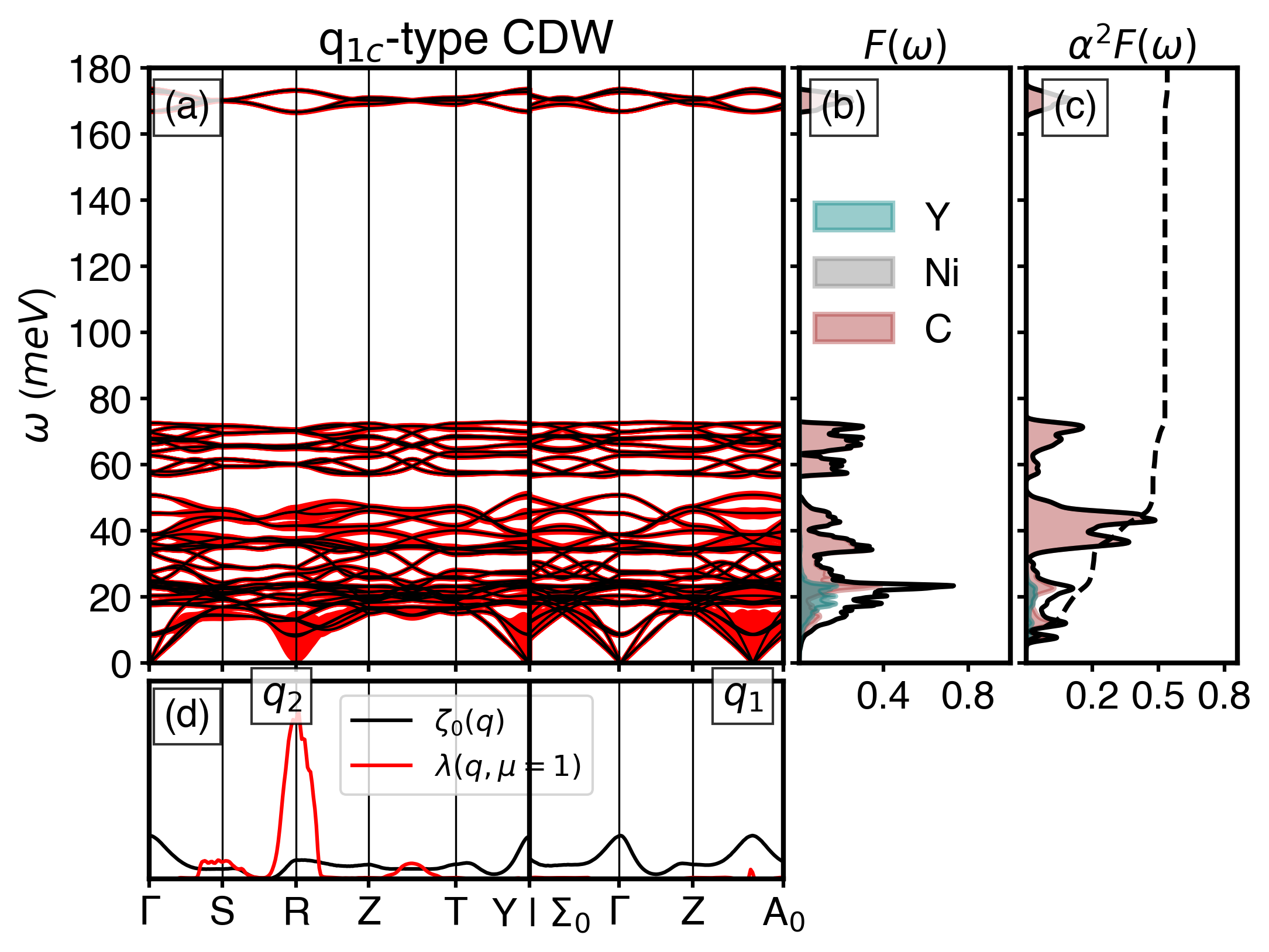}
\includegraphics[angle=0,width=0.52\columnwidth]{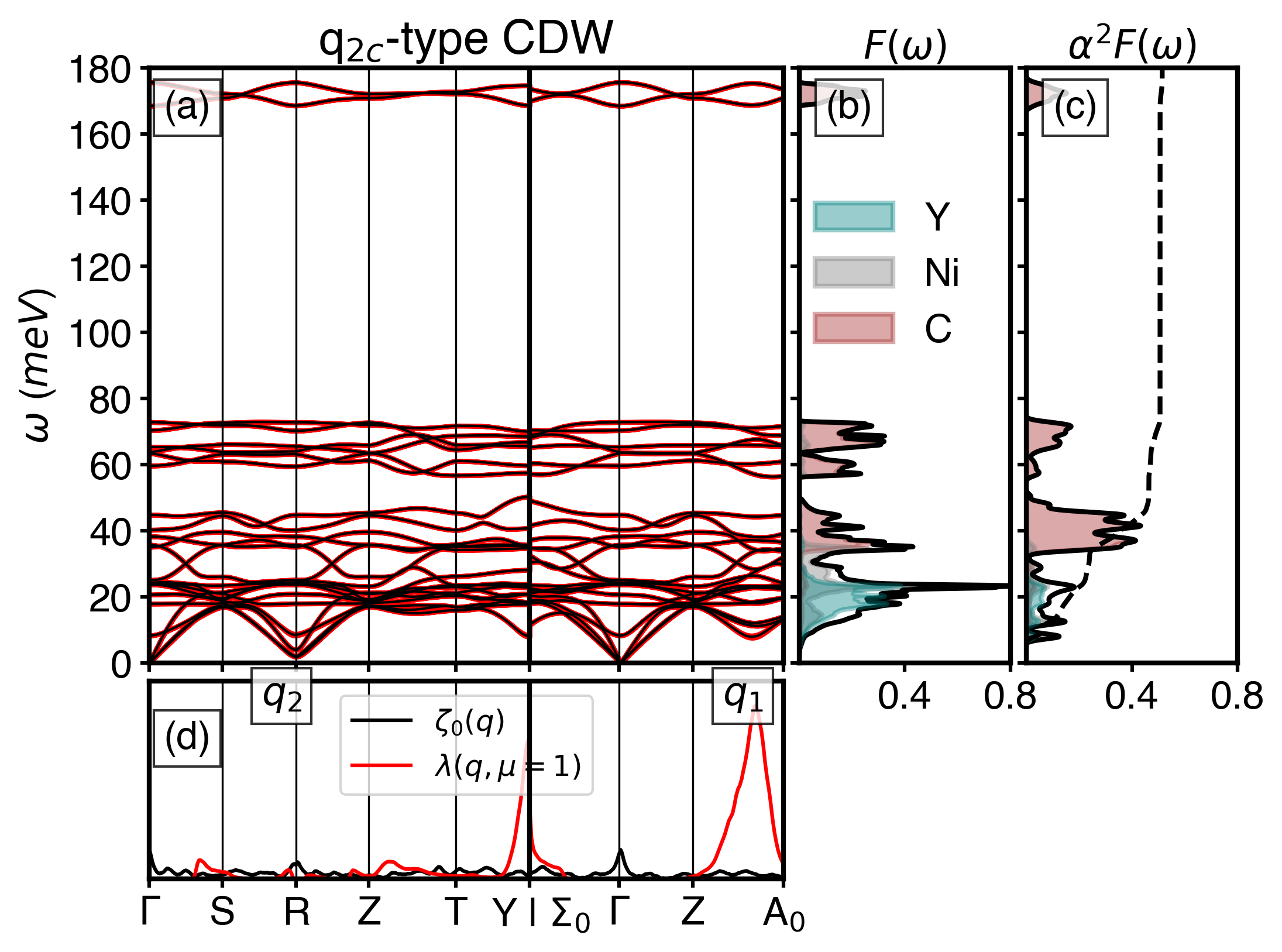}
\label{suppfig:phonons_q1q2}
 \caption{Phonon dispersion (a), atom-projected phonon density of states, $F(\omega)$, (b), Eliashberg function, $\alpha^2F(\omega)$, (c) and nesting function (d) for the $q_{1c}$-type (upper) and $q_{2c}$-type (lower) CDW phases. The projections onto Y, Ni, and C are indicated as yellow, grey, and red shaded areas, respectively.}
\end{figure}